\documentclass[fleqn,usenatbib]{mnras}

\usepackage{newtxtext,newtxmath}

\usepackage[T1]{fontenc}

\DeclareRobustCommand{\VAN}[3]{#2}
\let\VANthebibliography\thebibliography
\def\thebibliography{\DeclareRobustCommand{\VAN}[3]{##3}\VANthebibliography}

\usepackage{graphicx}	
\usepackage{amsmath}	

\title[Effects of stellar CMEs on atmospheric escape]{Magnetic interaction of stellar coronal mass ejections with close-in exoplanets: implication on planetary mass loss and Ly-$\alpha$ transits}
\author[Hazra et al.]{
Gopal Hazra,$^{1,2}$\thanks{E-mail:hazra@iitk.ac.in}
Aline A.~Vidotto,$^{2}$\thanks{Corresponding author: vidotto@strw.leidenuniv.nl}
Stephen Carolan,$^{3}$
Carolina Villarreal D'Angelo,$^{4}$
D{\'u}alta {{\'O} Fionnag{\'a}in}$^{5}$
\\
$^{1}$Dept. of Physics, Indian Institute of Technology, Kanpur, India\\
$^{2}$Leiden Observatory, Leiden University, PO Box 9513, 2300 RA Leiden, The Netherlands\\
$^{3}$School of Physics, Trinity College Dublin, Dublin 2, Ireland\\
$^{4}$Instituto de Astronom{\'i}a Te{\'o}rica y Experimental (IATE-CONICET), C{\'o}rdoba, Argentina\\
$^{5}$School of Mathematics, Statistics and Applied Mathematics, National University of Ireland Galway, University Road, Galway, Ireland}

\date{Accepted XXX. Received YYY; in original form ZZZ}

\pubyear{2024}

\begin{document}
\label{firstpage}
\pagerange{\pageref{firstpage}--\pageref{lastpage}}
\maketitle

\begin{abstract}
Coronal Mass Ejections (CMEs) erupting from the host star are expected to have effects on the atmospheric erosion processes of the orbiting planets. 
For planets with a magnetosphere, the embedded magnetic field in the CMEs is thought to be the most important parameter to affect planetary mass loss. In this work, we investigate the effect of different magnetic field structures of stellar CMEs on the atmosphere of a hot Jupiter with a dipolar magnetosphere. We use a time-dependent 3D radiative magnetohydrodynamics (MHD) atmospheric escape model that self-consistently models the outflow from hot Jupiters magnetosphere and its interaction with stellar CMEs. For our study, we consider three configurations of magnetic field embedded in stellar CMEs -- (a) northward $B_z$ component, (b) southward $B_z$ component, and (c) radial component. {We find that both the CMEs with northward $B_z$ component and southward $B_z$ component increase the planetary mass-loss rate when the CME arrives from the stellar side, with the mass-loss rate remaining  higher for the CME with northward $B_z$ component until it arrives at the opposite side.} The largest magnetopause is found for the CME with a southward $B_z$ component when the dipole and the CME magnetic field have the same direction. We also find that during the passage of a CME, the planetary magnetosphere goes through three distinct changes - (1) compressed magnetosphere, (2) enlarged magnetosphere, and (3) relaxed magnetosphere for all three considered CME configurations. We compute synthetic Ly-$\alpha$ transits at different times during the passage of the CMEs. The synthetic Ly-$\alpha$ transit absorption generally increases when the CME is in interaction with the planet for all three magnetic configurations. The maximum Ly-$\alpha$ absorption is found for the radial CME case when the magnetosphere is the most compressed. 
\end{abstract}




\section{Introduction}
Atmospheric escape from exoplanets is well observed in several discovered exoplanets, especially in the close-in exoplanets such as hot Jupiters and warm Neptunes \citep[e.g.,][]{Vidal-Madjar2003, Vidal-Madjar04,Lecavelier2010,Ehrenreich2015, Lavie2017}. The upper atmosphere of these planets gets photoionized due to the intense radiation from their host stars leading to an escaping planetary outflow \citep[e.g.,][]{Tian05, Murray-Clay2009, Allan2019, Hazra2020}. Continuous depletion of atmospheric material due to atmospheric escape over a long time is crucial for the sustainability and evolution of exoplanetary atmospheres \citep[e.g.,][]{Kubyshkina2020}. Also, atmospheric losses are important to understand the plausible cause of the existence of the Neptunian desert \citep[e.g.,][]{Mazeh2016} and the radius valley \citep[e.g.,][]{Fulton2017} in the present exoplanet demographic. A strong atmospheric mass-loss rate could even lead to the total loss of atmosphere\citep[e.g.,][]{Lammer2007, Khodachenko2007}. Therefore, for understanding the long-term evolution of the atmosphere, a more precise understanding of atmospheric escape and the corresponding mass-loss rate is necessary. 

Stellar radiation plays the key role in driving the planetary outflow from close-in exoplanets and once this radiation-driven planetary outflow starts to expand, it interacts with the stellar environment i.e., with the stellar wind, coronal mass ejections (CMEs), and stellar magnetic field. Presently there are some numerical efforts to understand mass loss due to the interaction of planetary outflow with stellar outflow using 3D hydrodynamic simulations \citep{Bisikalo2013, Villarreal2014, Villarreal2021, Tripathi2015, Carroll-Nellenback2017, McCann19,Carolan2020, Hazra2022, MacLeod2022}. However, limited studies are using 3D MHD simulations to understand these aspects. The stellar and planetary magnetic fields play a key role \citep{Khodachenko2015,Matsakos2015, Erkaev2017, Arakcheev2017,Daley-Yates2018, Daley-Yates2019, Zhilkin2020} in the interaction between the outflowing planetary material with stellar outflow. On one hand, the planetary magnetic field can suppress the atmospheric escape and change the morphology of planetary outflow depending on the structure of the planetary magnetosphere and its strength \citep{Trammell2014, Villarreal2018, Carolan2021, Khodachenko2021, Ben-Jaffel2022}. On the other hand, the stellar magnetic field reconnects with the planetary magnetic field and enhances the chance of planetary materials to leave the influence of planetary magnetosphere \citep{Lanza13, Owen2014, Egan2019, Ramstad2021}. The interplay between the stellar magnetic field and planetary magnetic field needs to be taken into account for a better understanding of the atmospheric escape from exoplanets \citep[e.g.,][] {Carolan2021}. 

Stellar transient activities (e.g., flares, CMEs) are also very important in affecting exoplanetary atmospheres. There are several studies on the effect of solar flares and CMEs on the atmospheres of solar system planets \citep[e.g.,][]{Ma2004, Manchester2004, Mostl2015, Falayi2018}. A few theoretical studies also investigated the effect of flares and CMEs on the exoplanetary atmosphere,  reporting that these stellar transients affect exoplanetary atmospheres greatly by changing the mass-loss rate \citep{Kay2016, Chadney2017, Bisikalo2018, Cherenkov2017,Hazra2020, Hazra2022}. A stellar flare enhances total X-ray and ultraviolet radiation received by the planet, potentially changing the atmospheric chemistry and ionization in the atmospheres of planets, which leads to more atmospheric escape \citep{Hazra2020, Louca2022} than when the star is in its quiescent phase.  Stellar CMEs enhance the stellar wind conditions, increasing the particle density and velocity of stellar wind as well as its embedded magnetic field, and hence, when a CME interacts with the planetary atmosphere, the planetary atmosphere gets disturbed significantly \citep{Cherenkov2017, Hazra2022}. 

Recently, \citet{Hazra2022} studied the effect of CMEs on the atmosphere of the hot Jupiter HD189733b and found that CMEs are very effective in stripping the planetary atmospheric material, increasing mass-loss rate and enhancing the transit signature in the Ly-$\alpha$ line. Similar observational enhancement of transit depth from visit to visit in Ly-$\alpha$ and helium lines is also reported due to the interaction of CME (strong stellar wind case) with the planetary atmosphere \citep{Rumenskikh2022}. However, \citet{Odert2020} modeled the interaction of CME with the planetary atmosphere for the same system (HD189733b) without any significant difference in the Ly-$\alpha$ transit. \citet{Cherenkov2017} also studied the effect of CMEs on hot Jupiters using a time-dependent simulation and reported an enhancement in the atmospheric mass-loss rate. Most of the previous studies used hydrodynamical (HD) simulation to capture the effect of CMEs on the exoplanetary atmospheres. However,  close-in gas giants (e.g., hot Jupiter and warm Neptunes) possibly have magnetic fields \citep{Cauley2019} and, because CMEs are highly magnetized plasma ejections from host stars, it is very important to consider the magnetohydrodynamic (MHD) interaction between a CME and planetary magnetosphere. 

In this paper, we investigate the effects of different magnetic field configurations embedded in CMEs and on the atmospheric escape from a hot Jupiter. We assume a dipolar planetary magnetosphere and vary the orientation of the embedded magnetic field in CMEs to understand how different CME magnetic structures will affect the planetary magnetosphere and corresponding mass-loss rate. We also predict the transit absorption in the Ly-$\alpha$ line during the passage of the different CMEs over the planetary atmosphere.

The structure of the paper is as follows. In the next Section, we discuss our 3D radiative MHD model and present the result of the quiescent case scenario where only the stellar wind is interacting with the planetary magnetosphere. In Section~\ref{sec:CME}, we explain different orientations of the magnetic field in CMEs and discuss their interaction with the planetary magnetosphere. The mass-loss rate for each considered case is also computed in this section. The synthetic Ly-$\alpha$ transit calculations and plausible observation of these predicted transit spectra are presented in Section~\ref{sec:transit}. Finally, we conclude our findings in Section~\ref{sec:conclusion}

\section{3D Radiation MHD Model}\label{sec:model}
We use the self-consistent 3D atmospheric escape model that was recently presented in \citet[for a magnetised scenario]{Carolan2021} and \citet[for an unmagnetised scenario]{Hazra2022}. Here we summarise those models and refer the reader to those works for their more detailed description. In our model, the photoionization due to
incident stellar radiation, collisional ionization, and corresponding planetary evaporation are calculated self-consistently. We adopt a similar simulation setup as studied in \citet{Hazra2022} for the HD189733 star-planet system but include the planetary magnetic field as a dipole \citep{Carolan2021}. The Cartesian box of the simulation domain has an extension of $x = [-20, +40]R_{\rm p}$, $y = [-40,+40]R_ {\rm p}$, and $z = [-32,+32]R_{\rm p}$ with the planet at the origin ($x = 0$, $y = 0$, $z = 0$), where $R_{\rm p}$ is the radius of the planet. The stellar radiation comes from the left side of the grid ($-x$). We solve 3D radiation magnetohydrodynamic equations in the rotating frame of the planet as follows:
\begin{equation}
\frac{\partial{\rho}}{\partial{t}} +\nabla \cdot \rho {\vec{ u}} = 0,
\end{equation}
\begin{eqnarray}\label{eqn:momentum}
\frac{\partial(\rho\vec{u})}{\partial t} + \nabla \cdot \Bigg[\rho \Vec{u} \Vec{u} + (P_T + \frac{B^2}{8\pi})I - \frac{\Vec{B}\Vec{B}}{4\pi}\Bigg] = \rho\bigg( \vec{g} -\frac{GM_{*}}{(r-a)^2} \hat{R}\bigg) ~\nonumber \\
- \rho (\vec{\Omega} \times (\vec{\Omega}\times\vec{R})-2(\vec{\Omega} \times \vec{u})), 
\end{eqnarray}
%
\begin{eqnarray}\label{eqn:energy}
\frac{\partial \epsilon}{\partial t} + \nabla \cdot \Bigg[\Vec{u}\Bigg(\epsilon + P_T + \frac{B^2}{8\pi}\Bigg)-\frac{(\Vec{u}\cdot\Vec{B})\Vec{B}}{4\pi}\Bigg] = \rho \bigg( \vec{g} -\frac{GM_{*}}{(r-a)^2} \hat{R}\bigg) \cdot \vec{u} ~\nonumber \\  - \rho (\vec{\Omega} \times (\vec{\Omega}\times\vec{R})) \cdot \vec{u} + {\cal H}-{\cal C}, ~~
\end{eqnarray}
%
\begin{equation}
    \frac{\partial\Vec{B}}{\partial t} + \nabla \cdot (\Vec{u}\Vec{B} - \Vec{B}\Vec{u}) = 0.
\end{equation}
Here $\Vec{u}$, $\rho$, $P_T$ and $\Vec{B}$ are the velocity, density, thermal pressure, and magnetic field respectively. $\epsilon$ is the energy density = $\frac{\rho u^2}{2} + \frac{P_T}{(\gamma -1)} + \frac{B^2}{8\pi}$. We assume the planetary atmosphere is purely hydrogen (neutral and ionized). $I$ is the identity matrix, $\gamma = 5/3$ is the adiabatic index. $M_\star$ is the mass of the host star and $\Omega$ is the orbital velocity of the planet. $\Vec{R}$, $\Vec{r}$ are the positional vectors in stellar and planetary frames respectively and $a$ is the orbital distance between the star and planet. Heating due to stellar radiation is incorporated in the term ${\cal H}$ and cooling due to emission of Ly-$\alpha$ radiation and collisional ionization is incorporated in the cooling term ${\cal C}$. In this model, the incident XUV stellar radiation is assumed to be plane parallel and concentrated at a monochromatic wavelength of frequency $\nu$ with an energy of 20 eV. The details of the radiation transfer prescription of stellar heating are given in equation~4 of \citet{Hazra2022}. We assume cooling is due to Ly-$\alpha$ radiation \citep{Osterbrock1989} and collisional ionization \citep{Black1981}.      

We also simultaneously solve two more equations along with MHD equations for tracking neutrals and ions
\begin{equation}
   \frac{\partial n_n}{\partial t} + \nabla \cdot n_n\vec{u} = \mathscr{R} - \mathscr{I},
\end{equation}
\begin{equation}
   \frac{\partial n_p}{\partial t} + \nabla \cdot n_p\vec{u} = \mathscr{I} - \mathscr{R},
\end{equation}
where $\mathscr{I}$ and $\mathscr{R}$ are the ionization rate (due to photoionization and collisional ionization) and recombination rate, respectively. The total ionization rate is 
\begin{equation}
 \mathscr{I} =\frac{ \sigma n_n F_{\rm xuv} e^{-\tau}}{{h\nu}} + 5.83 \times 10^{-11} n_e n_n \sqrt{T} \exp{({-1.578\times 10^5}/{T})}
\end{equation}
where, $F_{\rm xuv}$, $\sigma$ and $\tau$ are the incident XUV radiation flux, the cross-section of hydrogen for photo-ionisation and optical depth of the planetary atmospheres, respectively. $n_n$ is neutral density and $n_e$ is the electron density. T is the temperature of the gas. The recombination rate is
\begin{equation}
 \mathscr{R} = 2.7 \times 10^{-13} (10^4/T)^{0.9} n_e n_p,
\end{equation}
where $n_p$ is the ion density. $\mathscr{I}$ and $\mathscr{R}$ are given in cm$^{-3}$~s$^{-1}$. 

At the surface of the planet, the velocity of the planetary outflow is set as reflective (i.e., the velocity in the true and ghost cells are of the same magnitude but the opposite sign) as the inner boundary condition. The base neutral density, ionization fraction, and temperature are fixed as $4.0 \times 10^{-13}$ g~cm$^{-3}$, $10^{-5}$ and $1000$ K \citep{Hazra2022}. Initially, we fill all the cells of the simulation box with steady-state 1D planetary outflow described in \citet{Allan2019}. For the magnetic field, we have fixed the field strengths at $R = 0.5R_p$ such that the desired dipole strength is obtained at $R = 1R_p$. The dipole is fixed in the north-south direction aligned with the rotation axis of the planet (i.e., zero inclination angle). A floating boundary condition on the magnetic field is applied at the planet's surface where the gradient of the magnetic field is kept constant between true and ghost cells so that the field lines can respond to changes in the outflow. We use inflow limiting boundary conditions at the outer boundary except in the negative x boundary when the stellar wind is injected (for details see \citealt{McCann19}).

We adopt the same parameters as that of the HD189733 system. We consider that HD189733b has a dipolar magnetosphere with a surface, polar field strength of 10 G.
The stellar XUV radiation in our model enters the grid from the left side and in the quiescent phase (no flares and no CMEs) is calculated from the observed
X-ray luminosity of the star \citep{Lecavelier12}. The computed XUV flux at the orbital distance is F$_{\rm xuv}$ = $4.84 \times 10^4$ erg~cm$^{-2}$~s$^{-1}$ \citep{Hazra2022}. Our model is then able to self-consistently simulate the planetary outflow with this observed XUV stellar radiation and with the above-mentioned dipolar magnetosphere of the planet.  

As the planetary outflow interacts with the stellar wind in the realistic star-planet system, we inject a stellar wind from the left side of the grid where the host star resides and study its interaction with the radiation-driven planetary outflow. We follow a similar approach as described in \citet{Carolan2021} to inject the stellar wind. At the negative x boundary, we set a 
stellar wind velocity, temperature, density, and a magnetic field. These parameters are derived from a 1D polytropic model with a polytropic index 1.05, $T_{\rm wind} = 1.9 \times 10^6$ K and $\dot{M} = 3 \times 10^{-12} M_{\odot}/$yr for the host star HD189733A.
{Our choice of mass-loss rate was inspired by the assumptions adopted in the numerical simulations of \citet{Kavanagh2019}, albeit other model assumptions could lead to smaller values \citep{2022MNRAS.512.4556S}. Our adopted mass-loss rate is  150 times higher than the solar wind mass-loss rate of $\dot{M}_\odot = 2\times 10^{-14}~M_\odot~$yr$^{-1}$. It is indeed expected that stars more magnetically active than the Sun, like HD189733A, have mass-loss rates that are higher than that of solar wind \citep{wood2004, 2021LRSP...18....3V}. 
For example, the very active K-dwarf stars Speedy Mic and AB Dor have estimated mass-loss rates of $130$ and $350~\dot{M}_\odot$, respectively \citep{2019MNRAS.482.2853J}. Less active K-dwarfs like 70 Oph AB and 36 Oph AB have estimated mass-loss rates ranging from $7$ to $56 ~\dot{M}_\odot$ \citep{2021ApJ...915...37W}. Our chosen value of $150 ~\dot{M}_\odot$ fall within the aforementioned ranges.  
} 
The stellar wind magnetic field is assumed to be radial with a value of 2 G at the stellar surface \citep{Carolan2021}. The chosen parameters also make sure that the stellar wind at the orbital distance of the planet is super-Alfvenic, so that the interacting planetary outflow with stellar wind cannot travel upstream and affect the boundary condition. 

\begin{figure}
     \includegraphics[width=\columnwidth]{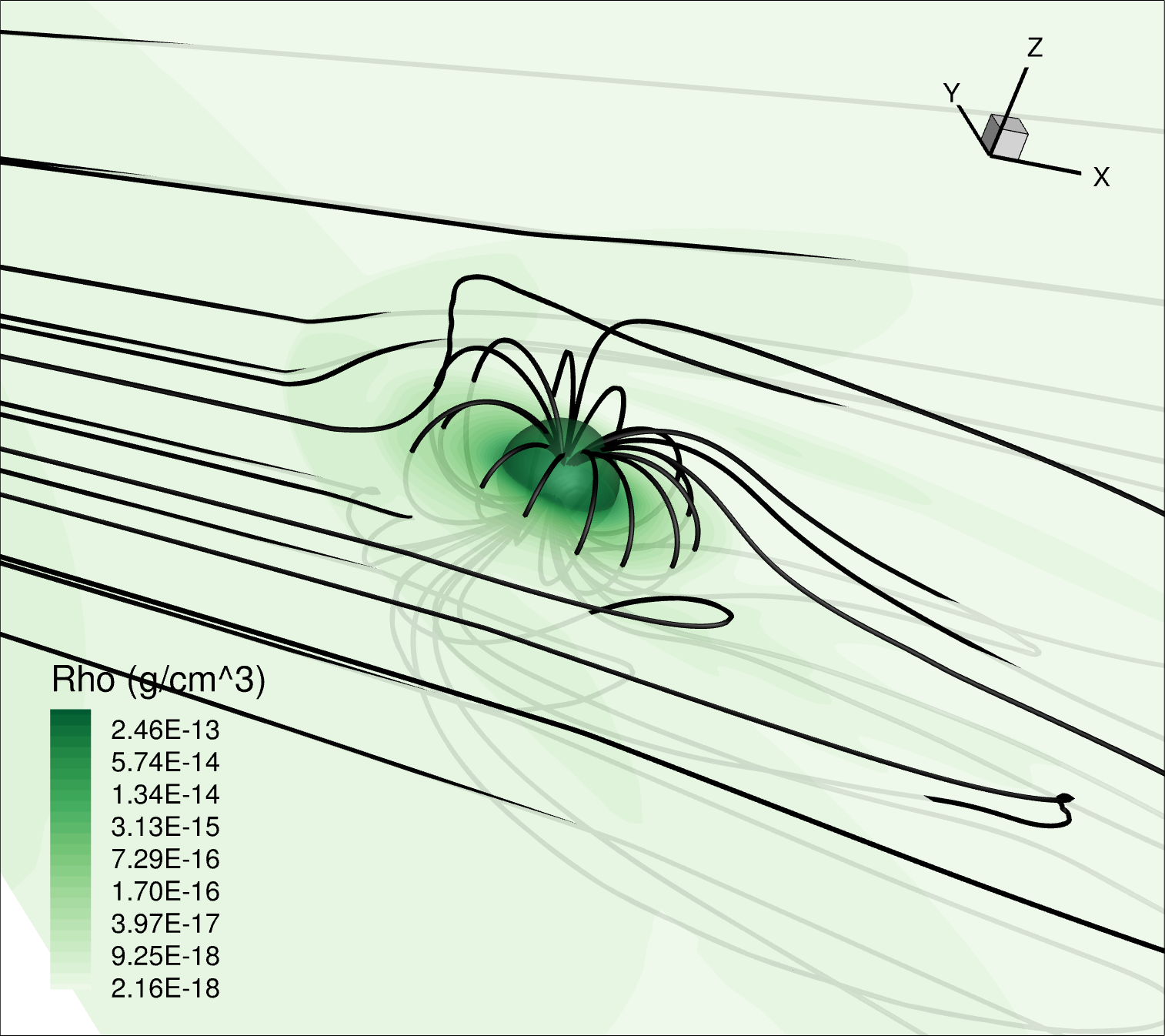}
    \caption{Total density of gaseous materials around the planet interacting with the stellar wind in quiescent phase. Black streamlines show the magnetic field lines. Stellar wind includes a radial magnetic field and the planet has a dipolar magnetosphere.}
    \label{fig:quiescent}
\end{figure}

The steady-state solution of the interacting stellar XUV radiation-driven planetary outflow with the stellar wind is given in Figure~\ref{fig:quiescent} during the quiescent phase. Here we have considered the quiescent phase of the star as a state with no flares and CMEs and estimated the XUV radiation at that state. The dipolar planetary magnetosphere is shown in black streamlines. The stellar wind interacts with the planetary magnetosphere and compresses the dayside magnetosphere. Compared to the hydrostatic case (section-3 in \citet{Hazra2022}), the presence of the planetary magnetic field changes the dynamics of the planetary outflow \citep{Carolan2021}. In the presence of a dipolar magnetosphere, the material gets trapped in the equatorial region, and funnels through the polar region. The trapped material in the equatorial region is visible in Figure~\ref{fig:quiescent}. {The mass-loss rate for this case is $5.9 \times 10^{10}$ g~s$^{-1}$.}
This result is similar to the case with the planetary magnetic field of 10 G presented in \citep{Carolan2021} {and comparable with the mass-loss rate estimation of existing other studies of HD189733b \citep[e.g.,][]{Guo2011, Salz2016, Odert2020, Rumenskikh2022, Hazra2022}}. 

Once we have simulated the steady-state solution of stellar wind interaction with the planetary outflow in the quiescent phase (as shown in Figure~\ref{fig:quiescent}), we are ready to inject the time-evolving CMEs to study their effect on the planetary atmospheres. \citet{Hazra2022} considered the result of the maximum interaction of CME with the planet as a steady state but a time-dependent model is required to fully understand the dynamic response of CME on the planetary atmosphere. Here we consider such a time-dependent model where real-time CME evolution is considered.

\section{Orientations of CME magnetic field and their effects on planetary atmosphere}\label{sec:CME}
Direct evidence of stellar CMEs affecting an exoplanetary atmosphere from observation is yet to come. However, \citet{Hazra2022} argued that the enhanced temporal variation observed for the hot Jupiter HD189733b during a Ly-$\alpha$ transit \citep[September 11 transit event][]{Lecavelier12} is most likely due to a CME erupting from the host star. {\citet{Rumenskikh2022} also found the enhancement of Ly-$\alpha$ transit depth for their simulation of CME and planet interaction (strong stellar wind case)}.  Among all of the cases (flare case, CME case and flare + CME case) that \citet{Hazra2022} considered, the CME only case was able to enhance the transit signature significantly but it was not able to explain the exact observed enhancement in transit depth. The interaction of a CME with the planetary atmospheres considered in \citet{Hazra2022} was hydrodynamic and stationary in the sense that only the maximum interaction of CME was considered. In this paper, we investigate the effect of the CME magnetic field on the atmosphere of the planet with a dipolar planetary magnetosphere using a time-dependent MHD model instead (section~\ref{sec:model}). 

As the CMEs are time-dependent phenomena that travel in the interplanetary medium and affect the planetary atmosphere, we need to first have an understanding of the duration of CME passing time over the planetary atmosphere. For solar CMEs, depending upon the speed of CMEs, the passage time takes several hours. For example, a CME event in December 2008 took around 18 hours to cross the planet Earth \citep{Mishra2013}. For stellar CMEs, observational estimates of arrival time and passage time are far beyond the current capability of our available instruments. As a result, we rely on the numerical understanding of stellar CMEs. The host star of our system is a K-dwarf of mass $0.82M_\odot$ with a rotation period of 12 days and no CME simulation for this star is readily available. For this reason, we adopt a simulated CME event for the K dwarf $\epsilon$-Eridani as reported in \citet{OFionnagain2022}. $\epsilon$-Eridani has the same mass ($0.82M_\odot$) and spectral type as our host star HD189733A  with a similar rotation period of 10.22 days and hence adopting a CME event from $\epsilon$-Eridani is a reasonably good approximation for HD189733A.    

\begin{figure*}
	\includegraphics[width=\textwidth]{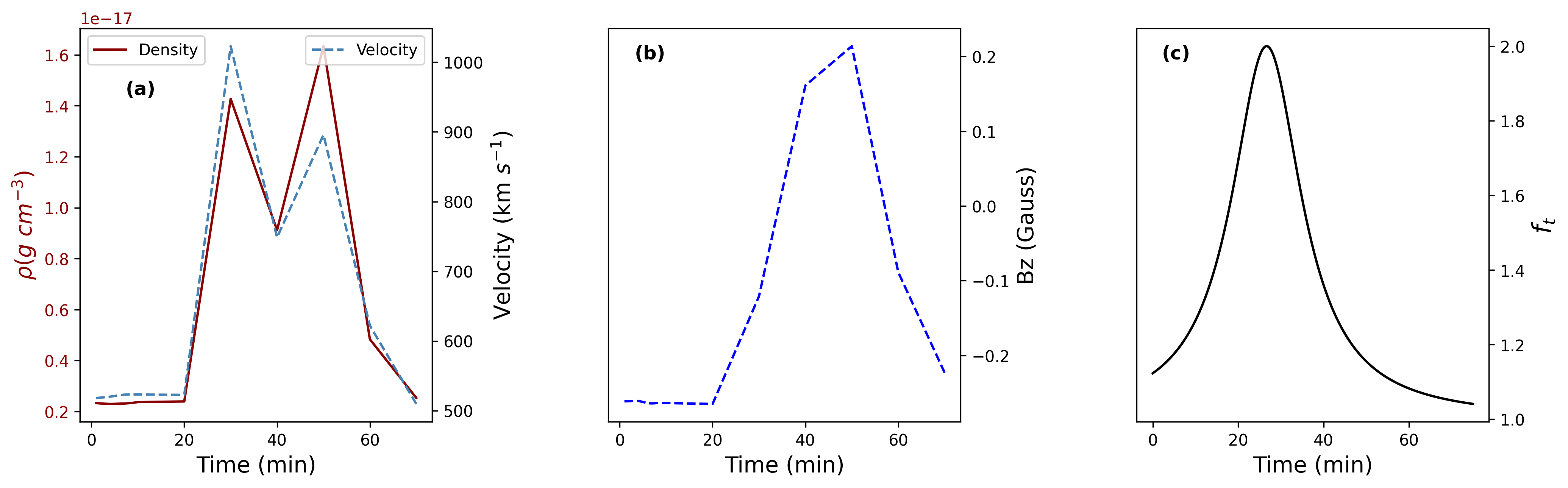}
    \caption{Evolution of simulated CME with time at a distance of 5.46R$_\star$ (0.02 AU) where the CME enters at the extreme left boundary of our simulation grid. (a) Total density (dark red) and velocity (blue) evolution. 
    (b) Time variation of the z component of magnetic field ($B_z$) embedded in the CME. (c) The time-evolving function used in our simulations to mimic the simulated CME time variation as shown in (a) and (b). The profiles presented in (a) and (b) are from the CME simulations of \citet{OFionnagain2022}.}
    \label{fig:cme_evols}
\end{figure*}

 We assume the CME as a spherical bubble with enhanced density, temperature, velocity, and magnetic field in comparison to the background stellar wind. The evolution of the considered CME in time is incorporated from the time-evolving equatorial CME (directed towards the planet) from \citet{OFionnagain2022}. This planet-effective CME has been simulated using an observed surface magnetogram from Zeeman Doppler Imaging (ZDI) on October 2013 for $\epsilon$ - Eridani (see bottom left of figure~5 for a snapshot of the CME solution at 10 minutes post-eruption in \citealt{OFionnagain2022}). 

%
The CME properties as a function of time (e.g., density, velocity, and temperature) in our simulation are chosen similar to the ones reported in \citet{OFionnagain2022} for the October 2013 case. Because in our grid the CME enters from the left boundary at a distance of $\sim 5.46 R_\star$ from the star ($20 R_p$ from the planet), the parameters of the CMEs extracted from \citet{OFionnagain2022} are taken at the same distance. Figure~\ref{fig:cme_evols}(a) and (b) show the evolution of the total velocity, density, and $B_z$ component of their simulation at this position. To include the time-dependence of each parameter in our model, we constructed a simple function $f_t$ (Figure~\ref{fig:cme_evols}(c)) with a different amplitude for each of the CME parameters to match the behavior of the CME properties found in \citet{OFionnagain2022}. We multiply the quiescent wind properties (density, velocity and temperature) by this function, assuming maximum amplitudes of  $f_t$  to be 8.0, 2.5, and 3.0 respectively. For example, the stellar wind density increases by up to a factor of 8 with respect to the quiescent wind density. The properties of our CMEs are very different from the ones considered in the study of \citet{Cherenkov2017}. The properties of their injected CMEs are based on solar CMEs, and they modelled them at three phases, with three different relative densities and velocities compared to their quiescent stellar wind. Their CMEs have maximum densities of a factor of 10 higher than the quiescent wind, while their maximum CME velocities are around 6 (slow), 13 (medium), and 30 (fast) times higher than the quiescent wind. 
%

Both the strength and geometry of the magnetic field in the CME are considered free parameters. In the context of the solar system planets, the geometry of the CME magnetic field plays an important role in contributing to a disturbance around planets with magnetic fields \citep[e.g.,][]{Wing1997, Falayi2018, Tenfjord2018}. If a CME carries a magnetic field in the southward direction (a negative $B_z$) relative to Earth's magnetosphere, it is found to be most effective in disturbing the magnetosphere and creating geomagnetic storms \citep{Nishida1983}. Following the understanding from the solar-system studies, we assume three geometric structures of the incoming CMEs in our present model as discussed below.

\subsection{Case-I: Northward $B_z$ only CME field}\label{sec:Bz_only}
We consider first a northward CME field which has only a positive $B_z$ component. The amplitude of 1 G for the magnetic field is specified in the left boundary ($-x$) of the grid where CME enters. The Mach Alfv\'en number (M$_{\rm Alfv}$) near the left boundary is 4.0 for the considered magnetic field strength in CME. The evolution of the CME parameters is included in the model using the time-dependent function explained in section ~\ref{sec:CME}.

\begin{figure*}
	\includegraphics[width=\textwidth]{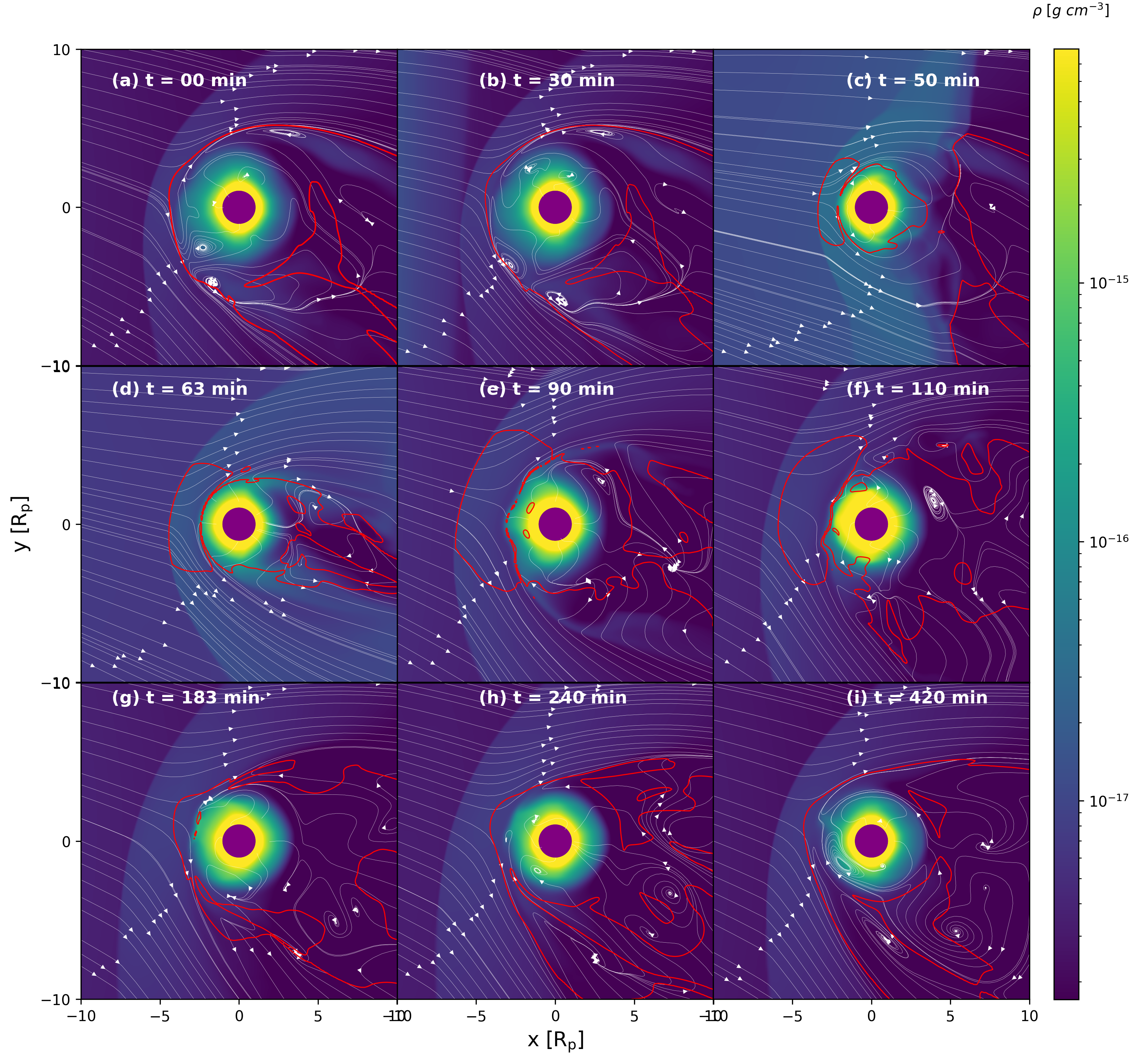}
    \caption{Total density of gaseous material around the planet for the Case-I CME interacting with the planetary atmosphere of the hot Jupiter HD189733b. Different snapshots show the scenario for different times (given in minutes) after the CME enters the simulation grid at $t=0$ min. White streamlines show the velocity field and the panels above show cuts in the orbital plane. The red contours show the Alfv\'enic surface ((M$_{\rm Alfv}$ = 1). }
    \label{fig:cme_bz_orbital}
\end{figure*}

The total density pattern of the planetary atmosphere in the orbital plane ($xy$ plane) is shown in Figure~\ref{fig:cme_bz_orbital}. Different snapshots in the figure show how the total density gets disturbed by the CME at different times. We mark the real-time in minutes (min) for each snapshot after CME enters at $t = 0$ min in the left boundary. The white streamlines show the velocity streamlines. The Alfv\'enic surface (M$_{\rm Alfv}$ = 1) is also shown for using the red contour line. At $t = 50$ min, the planetary magnetosphere reached its maximum compression on the dayside. At around $t = 63$ min, the CME interacts with the magnetotail, and the planetary magnetosphere becomes surrounded by the incoming CME. Eventually, after the CME has passed the planet, the planetary magnetosphere starts to get into a recovery phase (see Figure~\ref{fig:cme_bz_orbital}(f) at $t = 110$ min). It is only after 9 hours that the system gets back to its original pre-CME phase.  

\begin{figure*}
	\includegraphics[width=\textwidth]{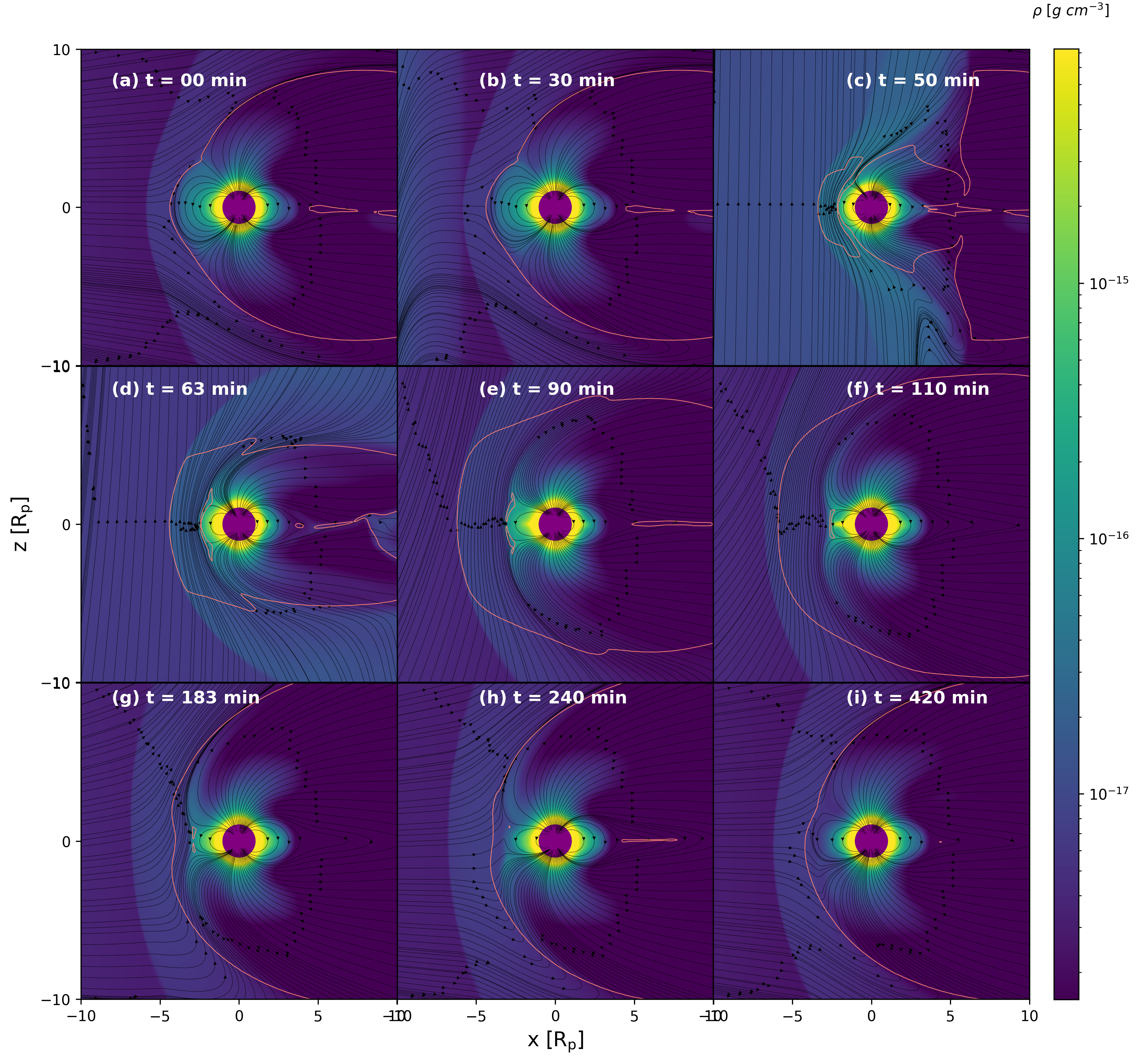}
    \caption{Density evolution of planetary material during the interaction of a CME with the planet HD189733b as seen in the polar plane ($xz$ plane). Black streamlines show the magnetic field lines representing the evolution of the magnetosphere during the interaction. Different snapshots are the same as figure~\ref{fig:cme_bz_orbital}. This case considers a CME  magnetic field of $B_z$ = +1 G (Case I). }
    \label{fig:cme_bz}
\end{figure*}

The magnetic field dynamics and the total density of the planetary atmosphere during the interaction of CME in the polar plane ($xz$ plane) are shown in Figure~\ref{fig:cme_bz}.
The dynamical response of the CME on the planetary atmosphere over time is clearly visible in this figure. Figure~\ref{fig:cme_bz}a shows the steady state condition of the interacting stellar wind with the planetary atmosphere before the CME enters the grid (pre-CME phase). The black streamlines show the overall structure of the magnetic field. Figure~\ref{fig:cme_bz}b shows the structure 30 min after the CME enters the simulation box, but note that it has not yet interacted with the planet's magnetosphere. Panel (c) shows the condition after $t = 50$ min where the CME already starts interacting with the planet suppressing the dayside magnetosphere due to the CME ram pressure ($\rho u^2$). Near the dayside, the z-component of the CME magnetic field is anti-parallel to the planetary dipolar magnetic field leading to magnetic reconnection, which allows some of the planetary material trapped in the equatorial dead-zone to escape. 

Figure~\ref{fig:cme_bz}(d) shows the situation after 63 min where the leading edge of the CME already crossed the planet but the planetary environment is still surrounded and disturbed by the CME plasma. This snapshot shows the situation of maximum disruption in the planetary environment overall. In the middle panel (Figure~\ref{fig:cme_bz}(e)), the CME is at the edge of the simulation box and the system is beginning to enter the recovery phase. Magnetic reconnection at the dayside still occurs and field lines start accumulating near the north and south polar sides. After 110 min (panel f), the CME has left the grid, and materials from the dayside are still outflowing instead of being trapped (figure~\ref{fig:cme_bz}(f)).

The third row shows the density pattern with the magnetic field streamlines in the post-CME situation. Panel (g) of Figure~\ref{fig:cme_bz} shows the situation after $t = 183$ min of CME eruption, where the incoming radial field in the embedded stellar wind starts to arrive on the planet diminishing the effect of CME $B_z$ field. Dayside outflow due to reconnection is still there with a large bow shock. The middle panel (figure~\ref{fig:cme_bz}(h)) shows that the planetary magnetosphere is coming back to its original form after $t = 240$ min of CME entered in the grid. However, despite a nearly similar overall pattern of the planetary magnetosphere, the dead-zone sizes are not the same as the case before CME enters. In the right panel (figure~\ref{fig:cme_bz}(i)), the dynamics after $t = 420$ min of CME eruption is shown. This figure shows that the planetary environment is trying to come back to normal phase after suffering from the CME disruption. Careful observation would show that this case is still not the same as the pre-CME phase and it takes another 180 min to get back to the original configuration.  

\subsection{Case-II: Southward negative $B_z$ only CME field}\label{sec:Bz_rev_only}
\begin{figure*}
	\includegraphics[width=\textwidth]{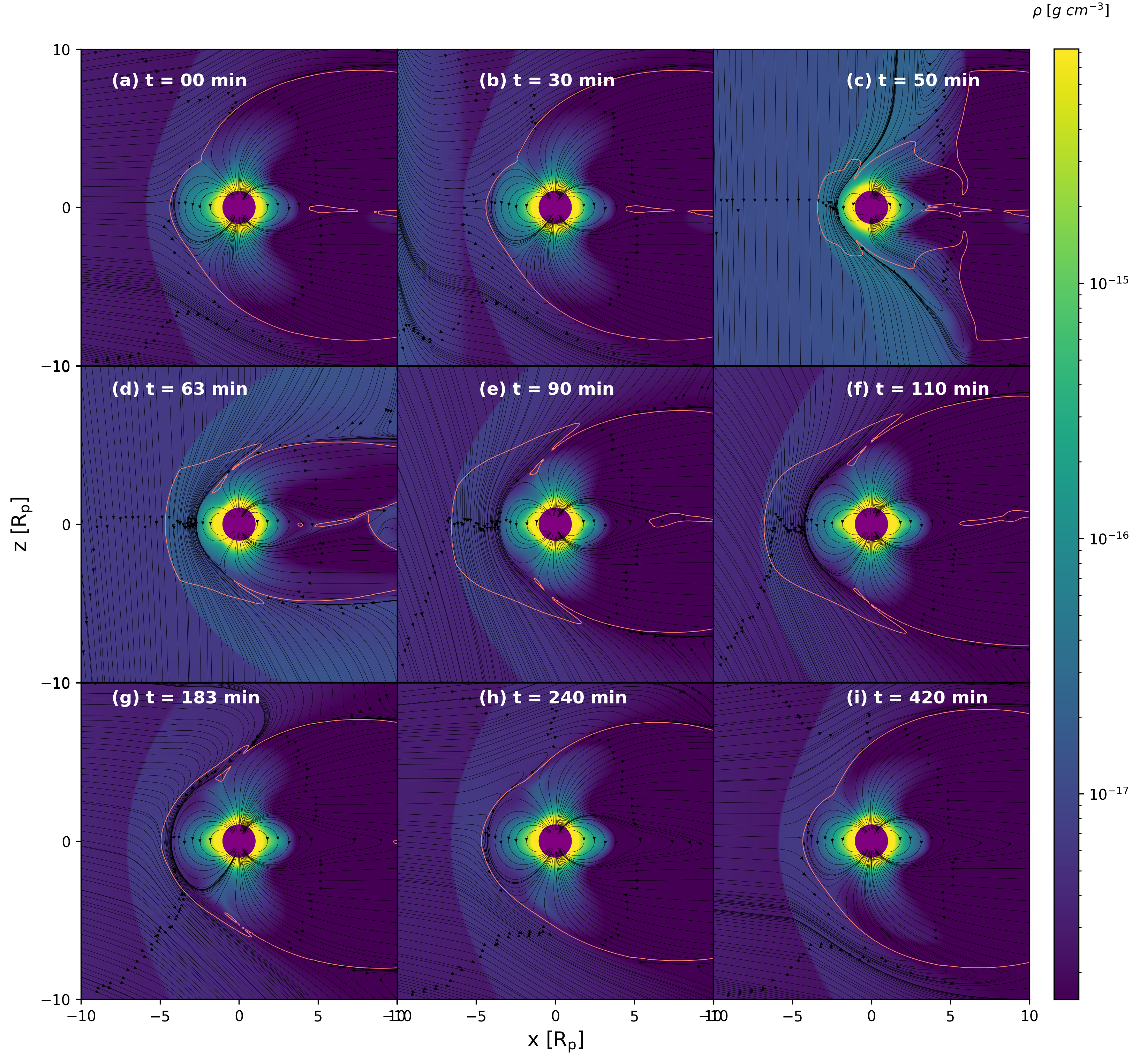}
    \caption{Same as the Figure~\ref{fig:cme_bz} but for the CME with a magnetic field $B_z = -1G$ (Case-II).}
    \label{fig:cme_bz_rev}
\end{figure*}

We also consider a CME with a southward magnetic field that is directed in the negative z direction. The amplitude of the magnetic field is taken the same as the Case-I in section~\ref{sec:Bz_only}. For this case, we only show the dynamics of the planetary atmosphere in the polar plane (xz plane) as it shows the most interesting and dynamic features. Snapshots of the total density pattern with magnetic field lines over time are shown in Figure~\ref{fig:cme_bz_rev}. Filled contours show the total density and black streamlines show the total magnetic field. The red contours show the Alfv\'enic surface. The times shown in this figure are the same as Figure~\ref{fig:cme_bz}. A major difference between the present case with the previous case of positive $B_z$ CME field is the different reconnection regions
due to the different orientations of the magnetic field. This is easily seen, for example, comparing panels d (maximum interaction 63 min after eruption) in Figures \ref{fig:cme_bz_rev} and \ref{fig:cme_bz}. 
When the leading edge of the CME is about to leave the simulation grid, 
we see magnetic reconnection now happening in the polar regions instead of the dayside magnetosphere as in the previous case (Case-I). This essentially changes the mass losses in comparison to the CME field with a positive $B_z$ component as seen in all the snapshots of figure~\ref{fig:cme_bz_rev}[(e)-(i)] and figure~\ref{fig:cme_bz}[(e)-(i)] over time.

\subsection{Case-III: Radial only CME field}\label{sec:CME_rad_only}
\begin{figure*}
	\includegraphics[width=\textwidth]{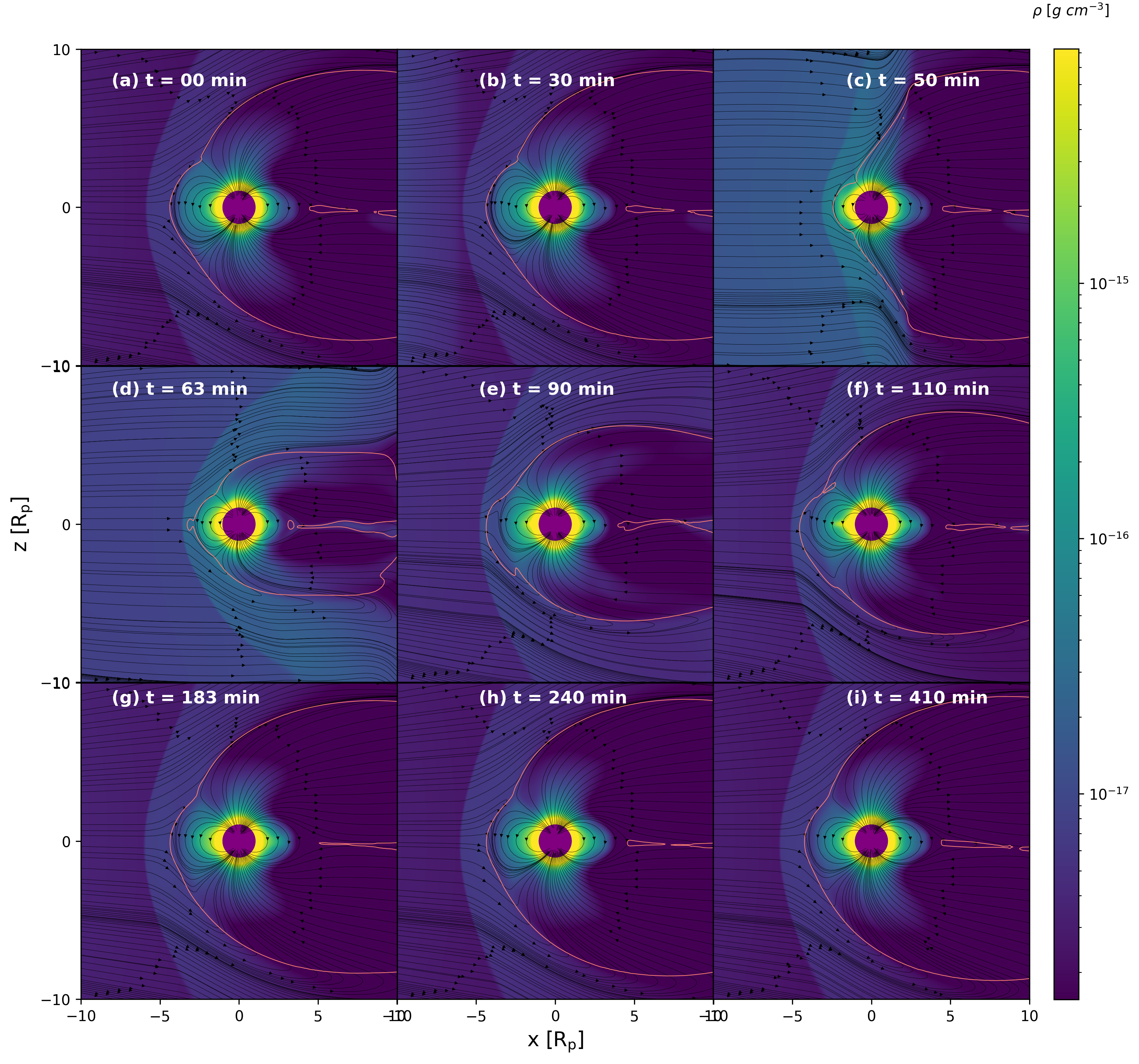}
    \caption{Same as Figure~\ref{fig:cme_bz} and \ref{fig:cme_bz_rev} but for CME field with radial component (Case III). Here the field strength is assumed to increase by a factor of 10, but the topology does not change with respect to the background stellar wind.}
    \label{fig:cme_radial}
\end{figure*}

In this case, we consider a CME which has a radial component of the magnetic field. This case essentially has the same geometry of the magnetic field as the background stellar wind but with enhanced field strength. Guided by the simulation of \citet{OFionnagain2022}, we have taken the x-component of the CME magnetic field 10 times the magnetic field strength for the background stellar wind. The snapshots of the time evolution of the interacting CME with the planetary atmosphere in the polar plane are given in Figure~\ref{fig:cme_radial}. In this case, the dynamics of the system are completely different in comparison to the CME with both z-components of the magnetic field (Case-I and Case-II). First of all, the enhanced strength of the magnetic field does not produce any significant changes in the overall magnetic geometry of the planetary magnetosphere. As there is no reconnection near the dayside or near the pole, the interaction with CME is mostly dominated by the dynamic pressure. The Alfvenic surfaces shown in red contours for all the snapshots support this. Also, the magnetosphere takes less time to recover in comparison to previous cases. The significant changes due to the consideration of different orientations of the CME magnetic field are prominent in the mass-loss rate calculation over time as we will see in the next section.   

\begin{figure*}
	\includegraphics[width=0.33\textwidth]{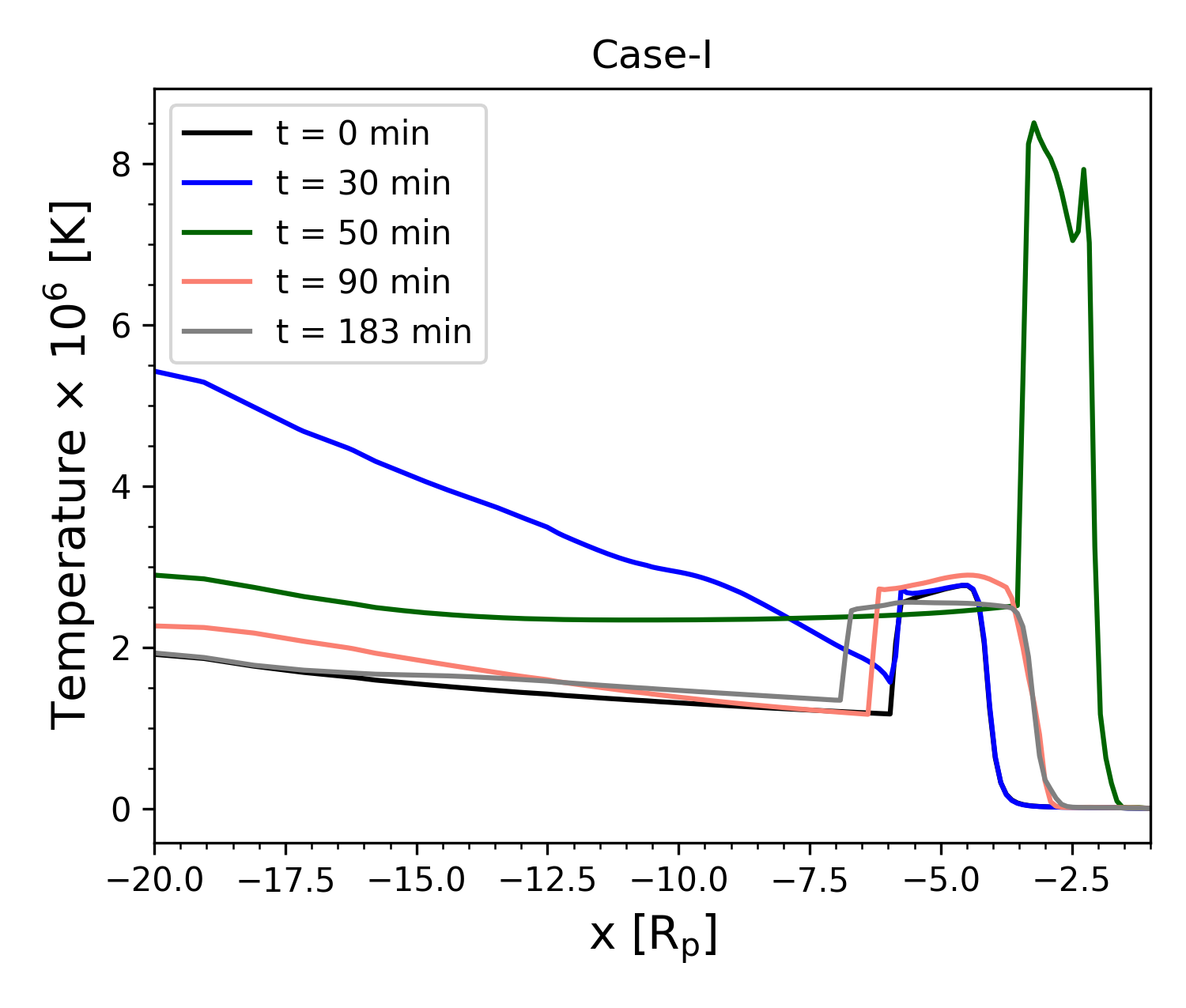}
     \includegraphics[width=0.33\textwidth]{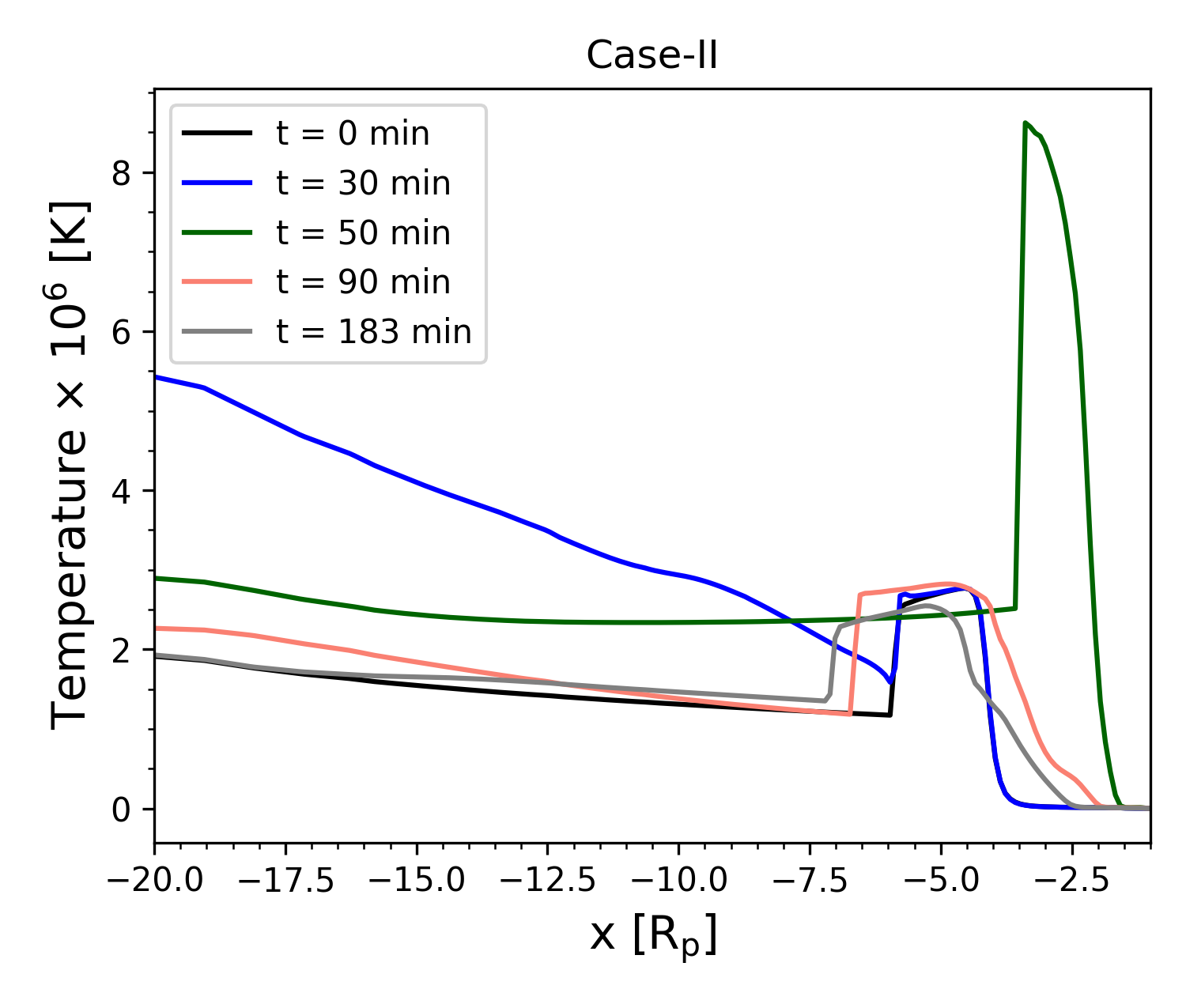}
     \includegraphics[width=0.33\textwidth]{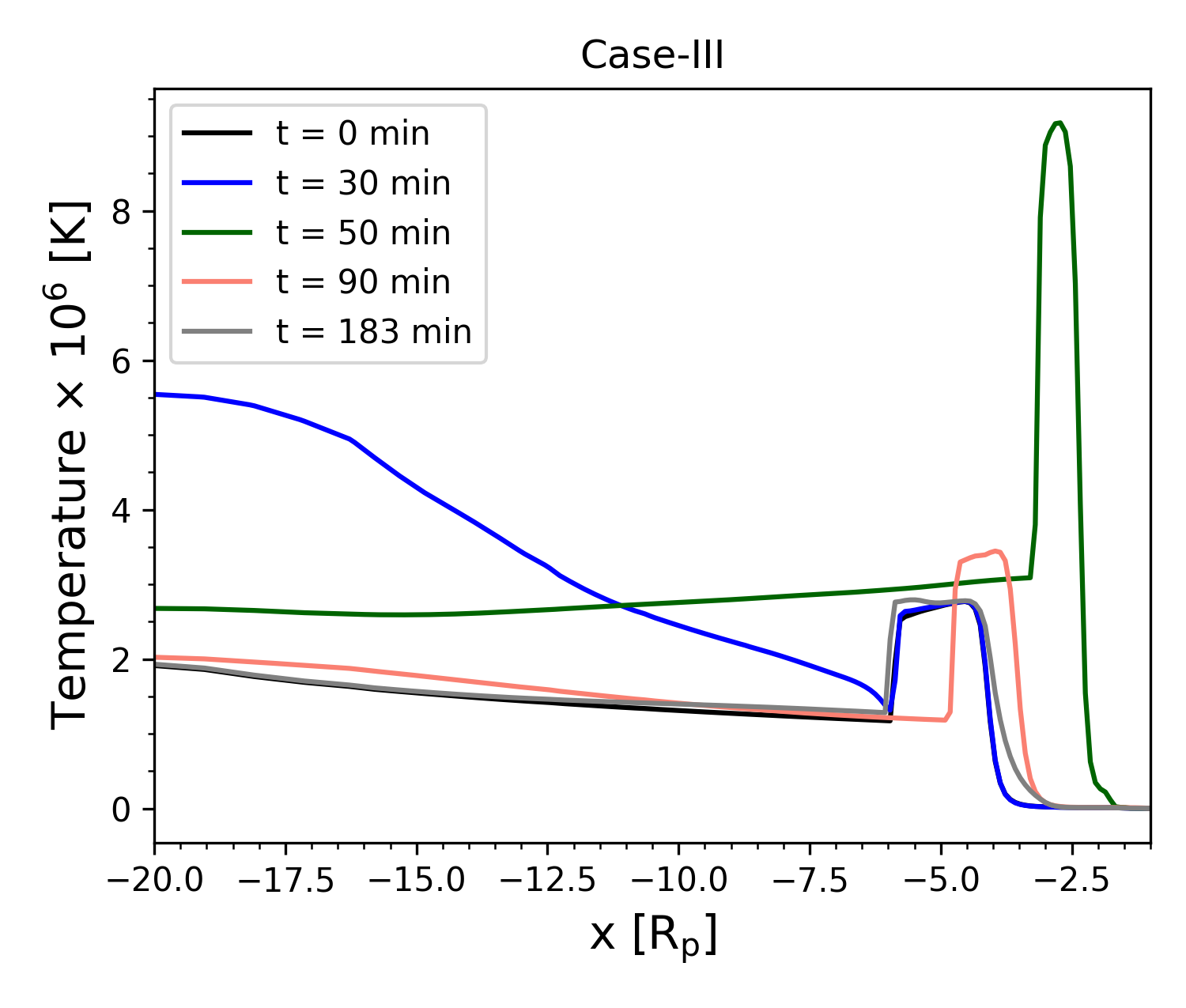}
    \caption{Variation of temperature along the star-planet line at different times for three different cases. The left, middle, and right panel shows Case-I, Case-II and Case-III respectively. The Black, blue, green, salmon and grey solid lines show the temperature variation at t = 0, 30 min, 50 min, 90 min, and 183 min after the eruption.}
    \label{fig:temp_starplanetline}
\end{figure*}

In Figure~\ref{fig:temp_starplanetline}, we show the temperature variation at five different times after the CME eruption for all the three considered magnetic geometries along the star-planet line in the substellar region. We show only five snapshots at five different times t = 0, t = 30 min, 50 min, 90 min and 183 min after the CME enters the grid, which are shown using black, blue, green, salmon and grey solid lines, respectively. The left, middle, and right panels of Figure~\ref{fig:temp_starplanetline} show Case-I, Case-II, and Case-III, respectively. The temperature variations at different times show that after CME hits the planet at 50 min producing a strong bow shock (green solid line), getting back to the pre-CME condition takes less time for the radial CME Case-III than the other two cases (compare the grey solid lines for three cases).

\subsection{Planetary mass-loss for different orientations of the magnetic field in the CME}
\begin{figure*}
	\includegraphics[width=\textwidth]{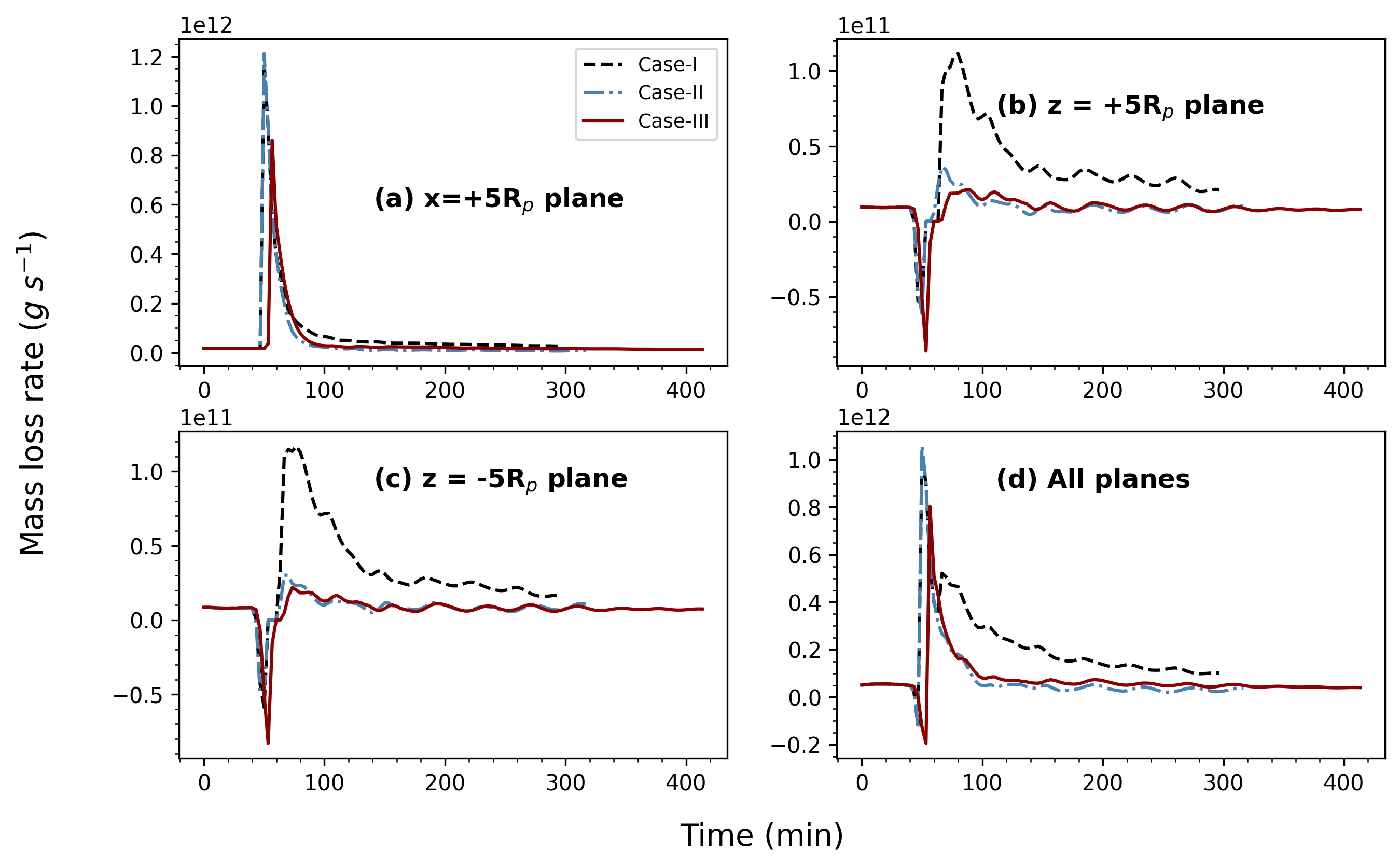}
    \caption{Mass-loss rate during the passage of CMEs from different planes around the planet - (a) from the $x= +5R_p$ at the night side of the planet, (b) from the $z=+5R_p$ plane at the north pole of the planet and (c) from the plane in the south pole at $z=-5R_p$. (d) The total mass-loss rate with time (i.e., summed over all six cubic planes). The x-axis shows the time in minutes after the CME enters the simulation grid. The black dashed and blue dash-dotted lines show the mass-loss rate for the CME with positive z-component (Case-I) and negative z-component (Case-II) of the magnetic field, respectively. The solid line in dark red shows the CME configuration with the radial magnetic field (Case-III).}
    \label{fig:mass_flux}
\end{figure*}

To get an estimate of mass loss from the planet over time, we calculate the mass-loss rate through different planes around the planet by integrating the mass flux. 
First, we consider the mass-loss rate from a $10 R_{\rm p}$ by $10 R_{\rm p}$ plane at $x= +5 R_{\rm p}$ in the planet's nightside. In Figure~\ref{fig:mass_flux}(a), the mass-loss rates through the $x= +5 R_{\rm p}$, are shown for the three considered magnetic structures of CME in our simulations. We see the mass-loss rates for both cases-I and II with $B_z = +1 G$ and $B_z = -1G$ can barely be distinguished (black dashed and blue dash-dotted lines respectively). There is a small reduction in mass-loss rate for the CME for Case-III with the radial B component, where we see that the planet is losing less mass through the $x =+5R_p$ surface compared to the other two cases.

The mass-loss rate through the north pole at $z=+5R_p$ is also shown in Figure~\ref{fig:mass_flux}(b). The mass-loss rate in this case is calculated by integrating the mass flux that passes through a plane of same area of $10 R_{p}$ by $10 R_p$ like the plane adopted before. For the $z=+5R_p$ plane, the mass-loss rate for each case is different over time. 
The mass-loss rate shows a momentarily negative dip around 50 minutes after the CME entered the simulation grid. This happens because the mass flux momentarily points towards the planet at this plane. Very quickly the escape process is resumed  and we see a positive recovery afterwards for all three cases. Integrated over time, we see that the effect of the CME is to increase atmospheric escape. 
The maximum mass-loss rate has been observed for Case-I with $B_z = +1 G$ just after the peak interaction phase when CME arrives at the opposite side. For the southern plane  at $z=-5R_p$ (also considering a plane with an area of $10 R_{p}$ by $10 R_{ p}$), the situation remains similar as shown in Figure~\ref{fig:mass_flux}(c).
The magnetic reconnection happens due to opposite alignment of CME magnetic field with planetary field for Case-I. As a result, the positive $B_z$ CME field shows more mass loss in comparison to the CME field with negative $B_z$. For both the northern and southern planes (figure~\ref{fig:mass_flux}(b) and (c)), the radial CME field (Case-III) shows a loss of the same amount of mass.  

Finally, Figure~\ref{fig:mass_flux}(d) shows the total summed mass-loss rate through all six planes of the cartesian cube centered on the planet. The volume of the cube is a sum of six planes each put at a distance of 5Rp from the planet. The planes at both positive and negative x , y, and z directions have the same area of $10R_p$ by $10 R_p$.
The total mass-loss rate is the same for both positive $B_z$ and negative $B_z$ components of the CME magnetic field before the peak interaction with the planetary atmosphere and it is easy to see that the main channel of escape is the night side by more than one order of magnitude. Once the CME arrives on the other side of the planet during the recovery phase, the mass-loss rate becomes higher for Case-1 with positive $B_z$. This is due to enhancements of mass-loss rates both at the north plane ($z=+5R_p$) and at the southern plane ($z=-5R_p$) for $t\gtrsim 60$~min. 
In summary, the CMEs with northward and southward $B_z$ components (Case-I and Case-II) are more effective in eroding the planetary atmosphere (stronger mass-loss) than the CME with radial field component (Case-III) in the early phase of CME impact. However, CME with northward $B_z$ component becomes more effective than the other two CME cases (Case-II \& Case-III) in eroding planetary atmosphere when CME arrives to the other side of the planet.

\subsection{Change in the planetary magnetosphere during the passage of a CME}
We investigate changes in the dayside magnetosphere when the CME passes through the planetary atmosphere from our simulations for the three cases. The traditional way to measure the extent of the magnetosphere is to calculate the magnetopause distance - the boundary that separates the shocked stellar wind plasma from the plasma inside the magnetosphere. We compute the dayside magnetopause as the distance from the planet, along the star-planet direction, at which the magnetic pressure of the planetary plasma is equal to the incoming CME/stellar wind ram pressure.   

\begin{figure}
	\includegraphics[width=0.45\textwidth]{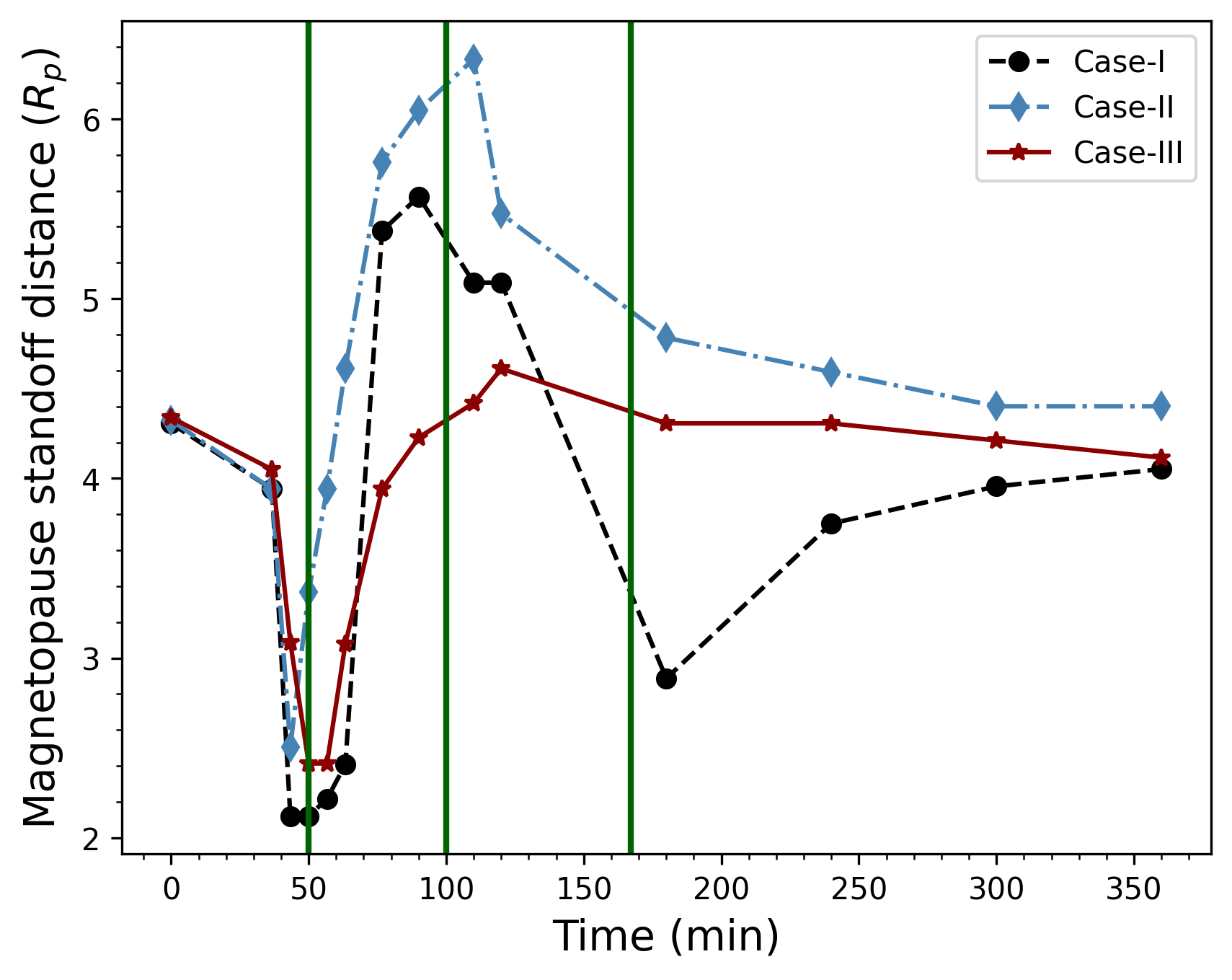} 
    \caption{Dayside magnetopause standoff distance as a function of time for three considered CME cases. The distances are given in units of planetary radius ($R_p$). The star symbol, filled black circle and blue diamond symbol show the magnetopause distance calculated for the radial CME field, positive $B_z$ CME field, and negative $B_z$ CME field respectively. The green solid vertical lines show three distinct phases of the planetary magnetosphere after the CME is injected into our simulation domain. We identify three distinct changes: (a) a compressed magnetosphere at $t \simeq 50$ min, (b) an enlarged magnetosphere at $t \simeq 100$ min and a relaxed magnetosphere at $t \simeq  167$ min.}
    \label{fig:magnetopause}
\end{figure}

In Figure~\ref{fig:magnetopause}, we plot the magnetopause distances for all of our three cases during the passage of the CMEs. The evolution of the magnetopause standoff distance with time for positive $B_z$ field (Case-I), negative $B_z$ field (Case-II), and radial CME field (Case-III) are shown using a filled black circle with the dashed line, a blue diamond symbol with dashed dot line and a star symbol with the solid dark red line, respectively. We identify that the planetary magnetosphere goes through three distinct changes when a CME passes through it. 

Around $t \simeq  50$ min after the CME injection in the simulation domain, the dayside magnetosphere gets compressed for all three kinds of incoming CME magnetic structures. Once the CME passes the planet at $\simeq 100$ min, 
the system starts to get back to its original form, and materials that were squeezed due to strong CME pressure flow back to the less-pressure zone, and hence the magnetosphere gets enlarged 
during this time for all three considered magnetic structures. The maximum magnetospheric distance is found for Case-II with negative B$_z$. This is because, for Case-II, no reconnection happens in the dayside as the southward CME field is in the same direction as the planetary magnetic field lines, resulting in the accumulation of planetary material within the enlarged magnetosphere. Eventually, at $t \simeq 167$ min, the system starts to go back to the relaxed configuration. These three distinctive phases of planetary magnetosphere -- compressed 
magnetosphere ($t \simeq 50$ min), enlarged magnetosphere($t \simeq 100$ min), and relaxed magnetosphere ($t =\simeq 167$ min) during the passage of a CME are marked with three solid green lines in the Figure~\ref{fig:magnetopause}.

\begin{figure}
	\includegraphics[width=0.48\textwidth]{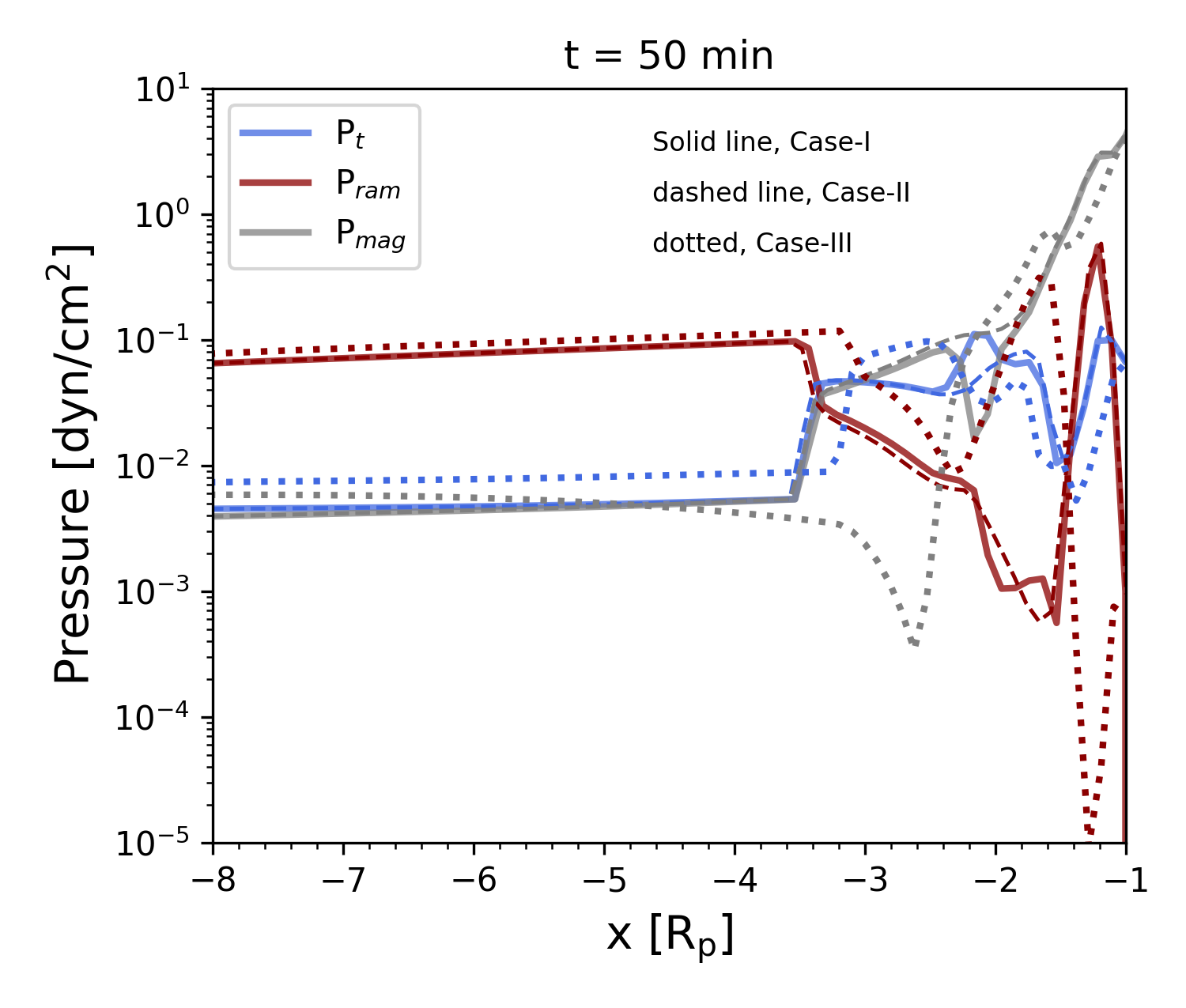} 
    \includegraphics[width=0.48\textwidth]{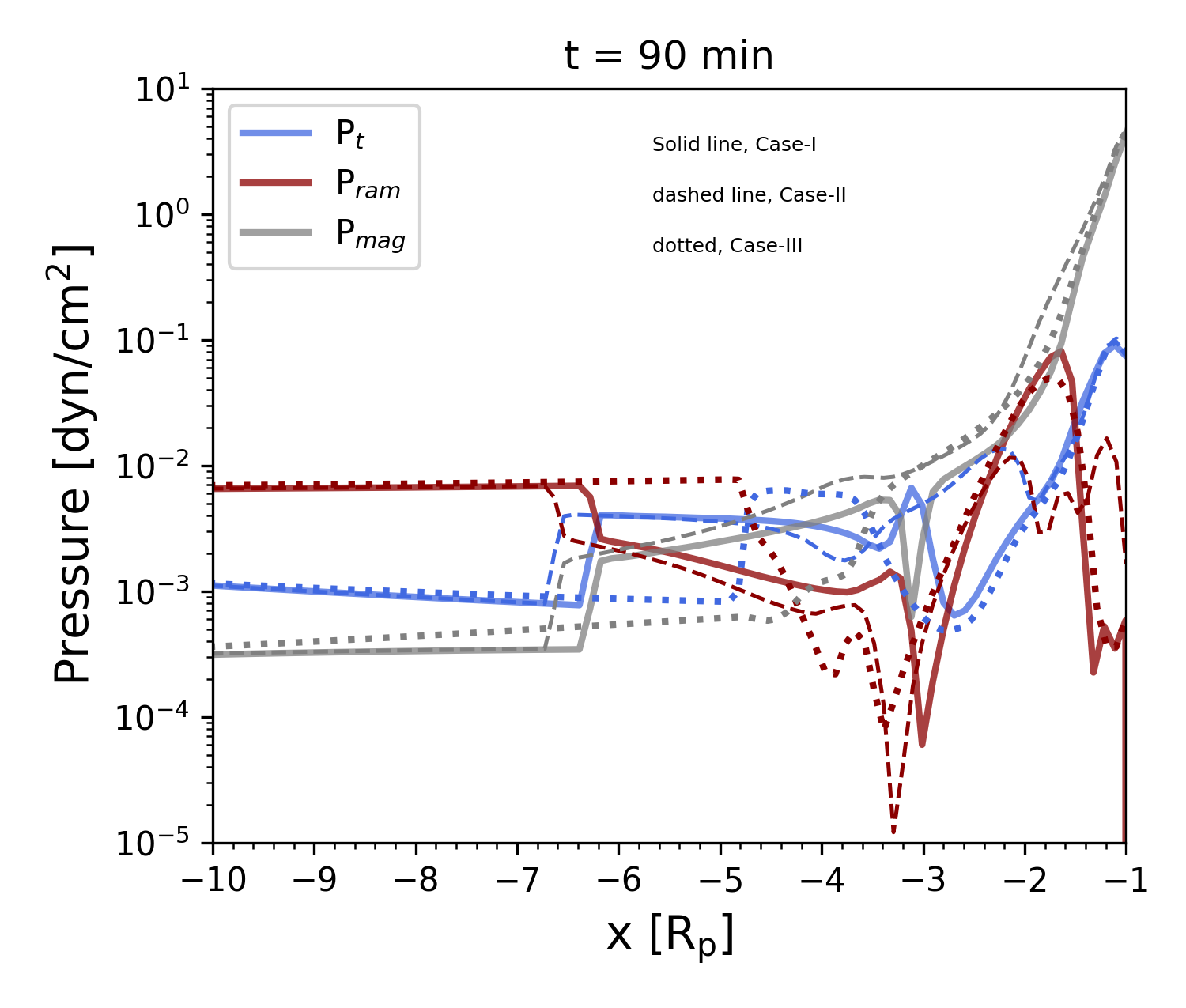}
    \caption{Upper Panel: the thermal pressure, ram pressure and magnetic pressure along star-planet line are shown using blue, dark red and grey colors at t = 50 min of CME eruption. These pressures for the positive $B_z$ CME field case (Case-II) are plotted using solid lines. The dashed and dotted lines show the other two cases with negative $B_z$ (Case-II) and radial component (Case-III). Lower Panel: Same as the upper panel but at t = 90 min after entering the simulation grid.}
    \label{fig:forcebalance}
\end{figure}

We have also plotted the thermal pressure, ram pressure, and magnetic pressure along the star-planet line in the dayside of the planet in Figure~\ref{fig:forcebalance} to understand region-wise force balance. The thermal pressure, ram pressure, and magnetic pressure are plotted using blue, dark red and grey colors respectively. All the three CME magnetic geometries of Case-I, Case-II and Case-III are plotted as solid, dashed and dotted lines respectively. In the upper panel of figure~\ref{fig:forcebalance}, we show the situation at t = 50 min after CME enters in our simulation grid. As expected, the ram pressure and magnetic pressure balance at the magnetopause, and hence below the magnetopause towards the planet, the stellar wind/CME ram pressure is overtaken by the magnetic pressure for all three cases. The situation remains the same  after t = 90 min of CME eruption (see bottom panel of Figure~\ref{fig:forcebalance}) but the magnetopause is moved further outside from the planet.

The different magnetopause standoff distances during the passage of a CME make it clear that the material bounded under the magnetosphere keeps changing when the CME interacts with the planetary atmosphere. As a result, when the planet is in transit, the transit signature would vary at different phases of CME interaction. In the next section, we calculate the synthetic transit signatures in the hydrogen Ly-$\alpha$ line during the different phases of CME interaction with the planetary atmosphere.


\section{Predicted Ly-$\alpha$ transit spectra during passage of a CME}\label{sec:transit}
In this section, we calculate the Ly-$\alpha$ transit depths for the three different magnetic geometries of the incoming CMEs at three different phases - (1) {\it Peak CME phase}: peak CME interaction phase during compressed magnetosphere at $t = 50$ min. (2) {\it CME passed phase}: phase just immediately after the CME left the planet at $t = 100$ min during the enlarged magnetosphere. (3) {\it CME relaxed phase:} recovery phase long after CME passed at $t = 167$ min. 
To calculate the transit spectra, we use a ray-tracing model used in previous works \citep{vidotto2018a, Allan2019, Carolan2020, Carolan2021a, Carolan2021, Hazra2022, Kubyshkina2022} and we refer the reader to them for more details. 
The atmospheric properties (e.g., temperature, neutral density, and velocity of the planetary neutral material) necessary to compute the spectra of the planet during CME events are taken from our radiation MHD simulation.

\begin{figure*}
	\includegraphics[width=\textwidth]{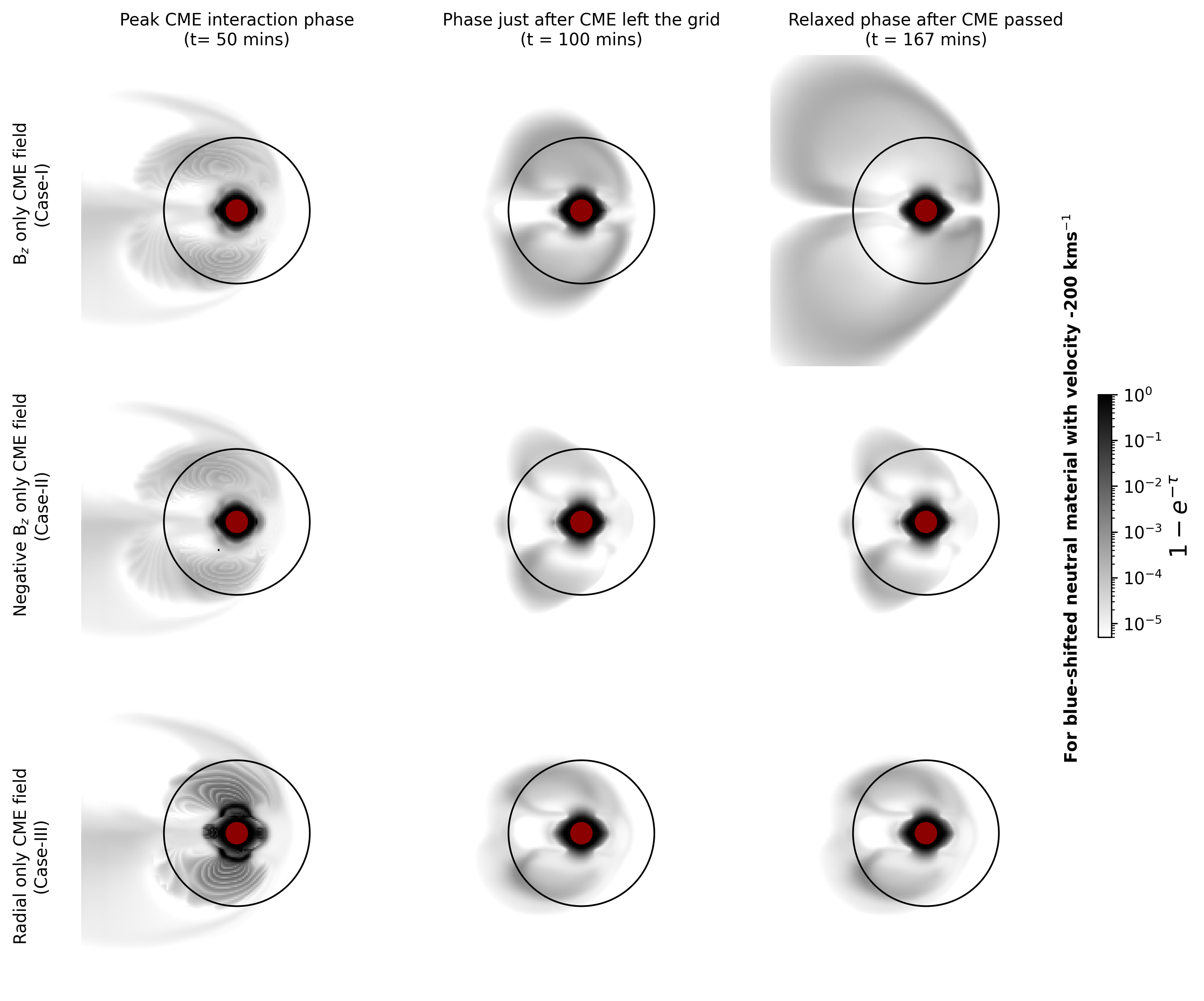}
    \caption{Line-of-sight opacity maps of blue-shifted Ly-$\alpha$ absorption at a velocity of -200 km~s$^{-1}$ for the three CME cases for three different interaction phases. The first row shows the Ly-$\alpha$ absorption during transit for $B_z$ only CME field (Case-I) at 50 min (CME peak phase), 100 min (CME passed phase), and 167 min (CME relaxed phase) at first, second and third columns respectively. The second and third rows plot the same for the CME field with negative $B_z$ only CME field (Case-II) and radial CME magnetic field (Case-III). The black circle shows the stellar disc and the filled red circle shows the planetary disc. }
    \label{fig:abs_bluewing}
\end{figure*}

In Figure~\ref{fig:abs_bluewing}, we show the absorption in Ly-$\alpha$ flux by the planetary neutral material in the plane of the sky. In this figure, we only show the opacity maps with a line-of-sight velocity of -200 km~s$^{-1}$ (blue wing), as the center of the Ly-$\alpha$ line [-40, +40] km~s$^{-1}$ is contaminated by the geocoronal emission and usually not considered in transit observations. The black circle represents the stellar disc and the red filled circle is the planetary disc.
The first, second, and third row of the Figure~\ref{fig:abs_bluewing} shows the Ly-$\alpha$ absorption by the planetary material for Case-I with positive $B_z$, Case-II with negative $B_z$ and Case-III with radial CME, respectively. The columns in the figure correspond to the {\it Peak CME phase}, {\it CME passed phase} and {\it CME relaxed phase}.
As the mass-loss rates are quite different for Case-I and Case-II during the late interaction phase when CME arrives on the night side of the planet, we see a very different distribution of material around the planet for these two cases (compare the opacity maps in the top and middle rows in Figure~\ref{fig:abs_bluewing}). As we will see next, these differences are not substantial and the Ly-$\alpha$ line profiles for Cases I and II remain approximately the same.

\begin{figure*}
	\includegraphics[width=0.45\textwidth]{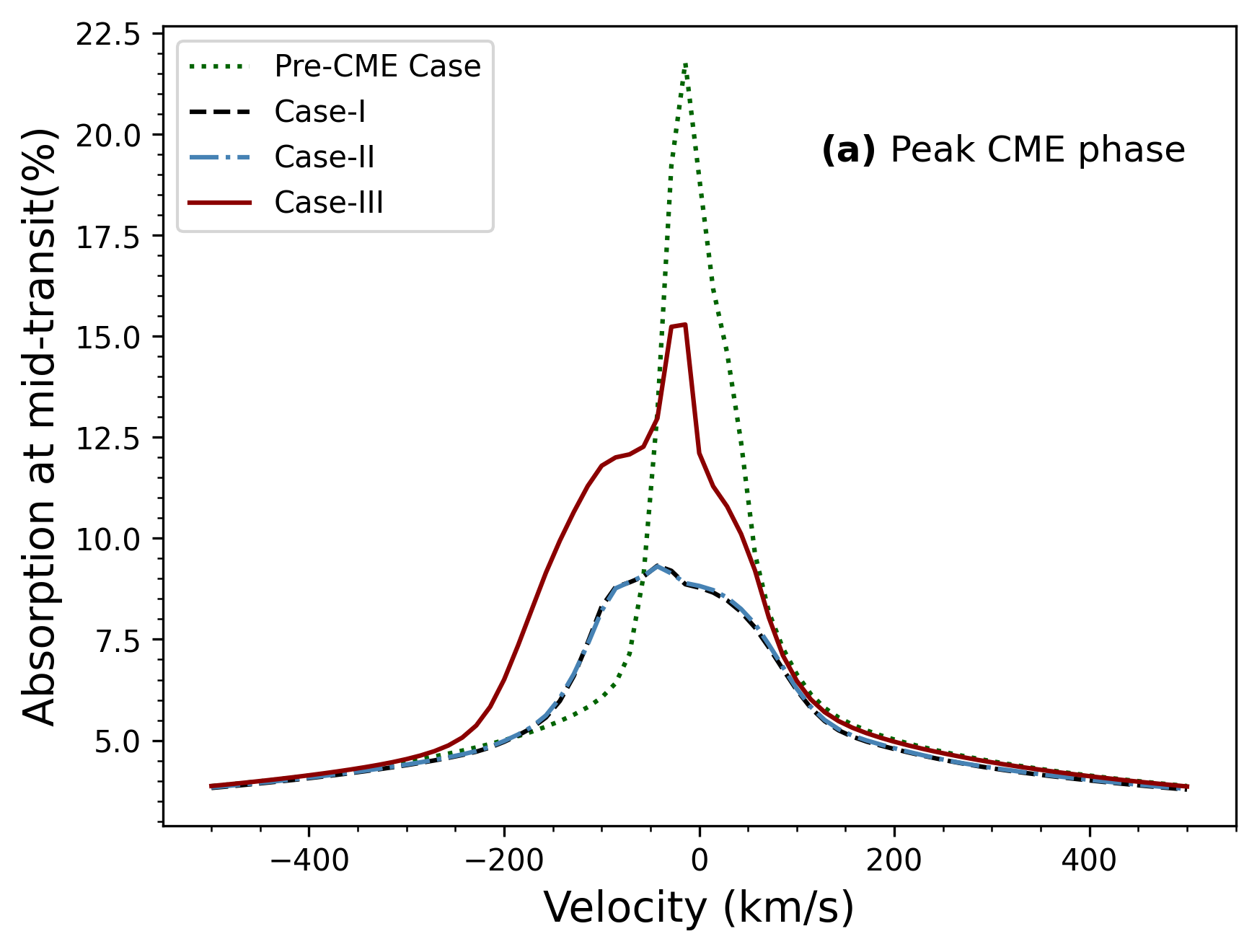}
     \includegraphics[width=0.45\textwidth]{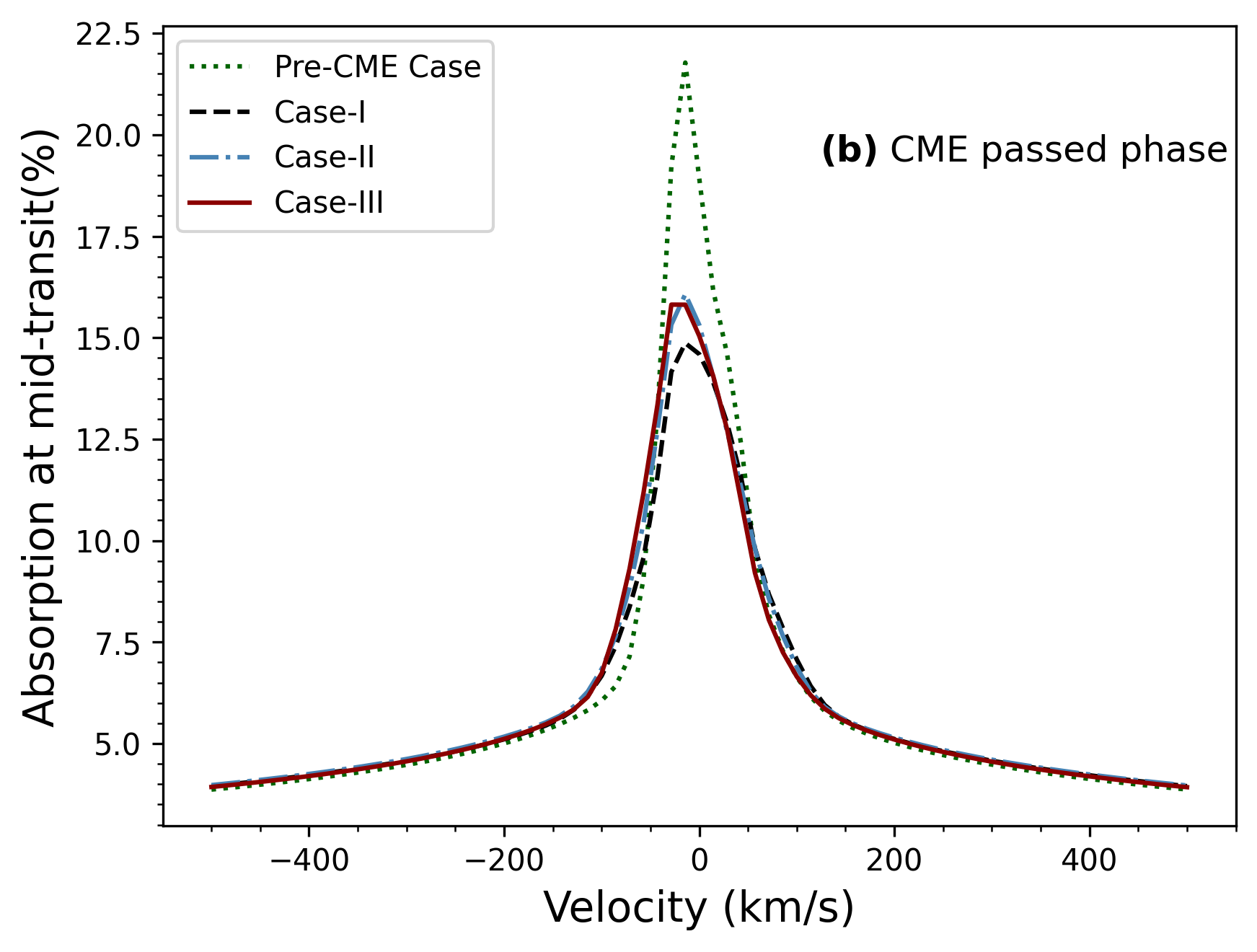}
     \includegraphics[width=0.45\textwidth]{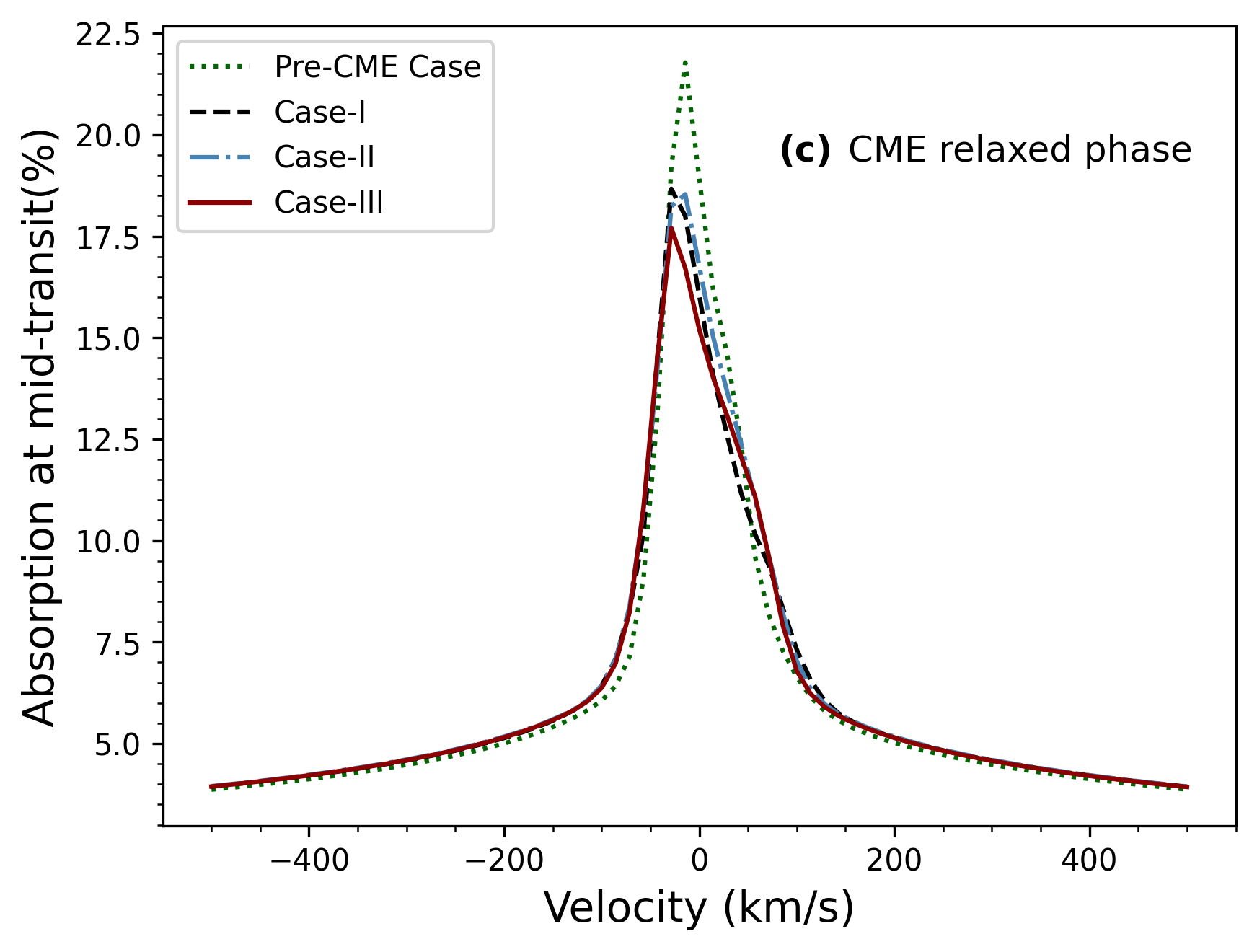}
     \includegraphics[width=0.49\textwidth]{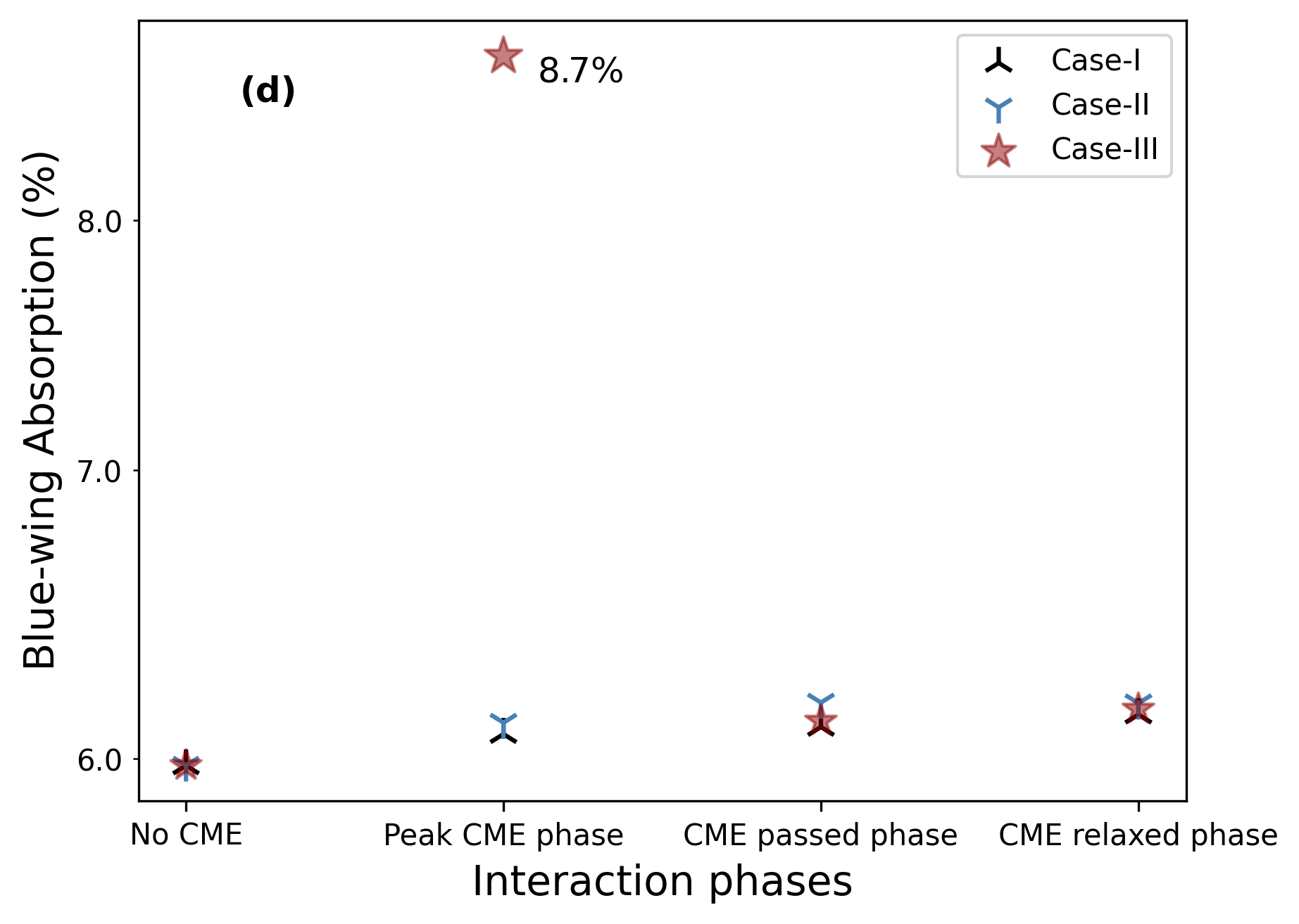}
    \caption{Ly-$\alpha$ line at mid-transit for different CME interaction phases for the three CME cases
    for the planet HD189733b with observed transit impact parameter = 0.6631. (a) Transit spectra at {\it Peak CME interaction phase} at time $t= 50$ min after CME eruption. The solid red, dashed black, and blue dash-dotted lines are the transit spectra for radial, positive $B_z$, and negative $B_z$ CME fields. The dotted green line shows the transit spectra for no CME case. (b) Same as (a) but for the {\it CME passed phase} at $t = 100$ min. (c) Transit spectra after $t = 167$ min at {\it CME relaxed phase}. (d) Scattered plot showing the integrated blue-wing transit absorption in the velocity range -230 to -140 km~s$^{-1}$ for all three CME cases at different interaction phases. Black upside triangle shapes, blue down triangle shapes, and red stars show the absorption depth for cases with $B_z$ and negative $B_z$ and radial CME field respectively.}
    \label{fig:Ly_alpha_b=0.6631}
\end{figure*}

To compare with the observed temporal variation of Ly-$\alpha$ transit spectra of HD189733b \citep{Lecavelier12}, we compute the synthetic transit spectra considering the observed impact parameter $b = 0.6631$ for three CME cases at different times during the CME passage (note that in the opacity maps shown in Figure \ref{fig:abs_bluewing}, we show the planet transiting at mid-disc).  Figures~\ref{fig:Ly_alpha_b=0.6631}(a), (b) and (c) show the Ly-$\alpha$ line at mid-transit for three phases of CME interaction phases. Transit spectra for each case are plotted using different colors. We also plot the transit spectra in the case when the planet is not experiencing any CME event (pre-CME case) to compare how the CME changes the transit depth.
In figure~\ref{fig:Ly_alpha_b=0.6631}(d), we show the total blue-wing integrated Ly-$\alpha$ flux absorbed in the velocity range -230 to -140 km~s$^{-1}$ for all CME cases for the three interaction phases (including pre-CME/no CME case) of the CME event. 

It is clear from Figure~\ref{fig:Ly_alpha_b=0.6631} that the CME event during different phases of interaction enhances the transit depth of the observable blue-wing in the velocity range -230 to -140 km~s$^{-1}$ compared to the no CME event. The CME suppresses the dayside magnetosphere more in comparison to the stellar wind (pre-CME phase), and Ly-$\alpha$ absorption in the line-center decreases during the CME event. 
The blue-wing absorption due to the high velocity materials increases in the CME event because the CME drags the high velocity neutrals with it. Overall the CME shifts the neutrals in the atmosphere towards the tail increasing their velocity. Depending on the geometry of the incoming CME magnetic field, the dynamics of the planetary material changes, resulting in different blue-wing absorption during different phases of the CME interaction as seen clearly in the figure~\ref{fig:Ly_alpha_b=0.6631}(d).

Contrary to Case-I and Case-II, Case-III does not encounter any magnetic reconnection (as the magnetic field line is the same as the incoming stellar wind) and the planetary dynamics due to CME interaction is mainly resulting from the CME ram pressure. 
During the peak CME interaction phase ($t = 50$ min), the dayside magnetosphere gets compressed due to incoming CME (see red solid line in Figure~\ref{fig:magnetopause}) but the planetary material at the night side follows the CME material producing high-velocity neutrals. As a result, we get enhanced transit depth in the blue-wing for the CME with radial magnetic field (see red solid line in Figure~\ref{fig:Ly_alpha_b=0.6631}(a)). For the CME with positive $B_z$, dayside reconnection happens with the planetary magnetosphere (see Figure~\ref{fig:cme_bz} and the equatorial dead-zone gets disturbed enhancing the blue-wing transit absorption. However, for the CME with negative $B_z$ magnetic field, no dayside reconnection is seen but there are reconnections in the polar regions (see Figure~\ref{fig:cme_bz_rev}). As a whole, it shows similar blue-wing absorption as the positive $B_z$ CME case. 

During the phase when the CME just passed the planet at $t = 100$ min (CME passed phase), we do not see significant differences in the blue-wing absorption for the different magnetic geometries. 
At $t = 167$ min after CME eruption in the CME recovery phase, the overall transit absorption increases without significant differences among the three considered CME cases. Note that, in our models, we do not consider charge-exchange reactions that could generate Energetic Neutral Atoms (ENAs). These ENAs could enhance the observed Ly-$\alpha$ transit depth during the interaction between CME and planetary outflow \citep{Khodachenko2019, Rumenskikh2022}. 
However, these charge-exchange reactions require careful modeling. For example, \citet{Odert2020} could not find a significant production of ENAs in their simulation of CME and planetary outflow interaction, because of the absence of a planetary tail where most of the ENAs are generated. 
Additionally, some of the recent simulations reported that the ENAs due to charge exchange do not significantly enhance the transit dept of Ly-$\alpha$ \citep{Esquivel2019}. We plan to investigate the effect of ENAs during CME interaction with the planetary atmosphere in a future work.

\section{Conclusions}\label{sec:conclusion}
In this paper, we simulated the interaction of a coronal mass ejection (CME) with a magnetized hot Jupiter atmosphere, considering the HD189733 system for this study. As  CMEs are time-dependent phenomena, they all propagate and interact with the planetary atmosphere within a few hours after the eruption, and hence all of our simulations were performed in time-accurate mode, i.e., in real-time after the CME has entered the simulation grid.  We use the 3D radiation MHD model developed in \citet{Hazra2022} and \citet{Carolan2021} to implement the time-accurate simulation of CME interaction with the planetary atmosphere. 
The properties of our injected CME were extracted from MHD simulation of $\epsilon$ Eri, a K-type star that has the same rotation rate and mass as HD189733A \citep{OFionnagain2022}. We kept the geometry of the embedded CME magnetic field as a free parameter and studied the effect of different geometries on the atmospheric escape and the corresponding transit signatures in the Lyman-$\alpha$ line.

The density, velocity, and temperature of the CME are considered to reach up to 8.0, 2.5, and 3 times those of the background stellar wind respectively, approximately reproducing the enhancement that each quantity follows in the simulated CME from \citet{OFionnagain2022} with respect to the background stellar wind. 
We considered three cases for the geometry of embedded magnetic field in the CME - Case-I: northward $B_z$, Case-II: southward negative $B_z$ and Case-III: radial field. For all three cases, the CME interacts with the planetary magnetosphere enhancing the planetary mass loss during the CME interaction time. The mass loss from the planet starts to increase as soon as the CME reaches
the planetary atmosphere, it becomes maximum when the CME is mid-way of crossing the planet and eventually decreases when the CME has passed the planet. The planetary atmosphere reinstates the original state after a few hours of complete passing of the CME. Depending upon the geometry of the incoming CME magnetic field, the system experiences different magnetic reconnection and planetary mass-loss rate differs. CMEs with $B_z$ only component - both positive (northward) and negative (southward) are more effective than the CME with the radial magnetic field in removing planetary material from the planet during the early phase of the interaction when CME enters the stellar side. Later on, when CME arrives on the opposite side of the planet, CME with positive (northward) $B_z$ becomes more effective in removing planetary materials with a higher mass-loss rate.  

The size of the planetary magnetosphere changes when a CME passes through it. We have calculated the dayside magnetopause-standoff distance for all three CME cases. We found that at the maximum CME interaction time, the magnetopause gets reduced (compressed magnetosphere). When CME crosses the planet, the system starts to get back to its original form, and materials that were squeezed due to strong CME pressure flow back to the less-pressure zone, and hence the magnetosphere gets enlarged. Eventually, after a few hours, the system recovers and the magnetosphere moves to the pre-CME original size. Among all three considered CMEs, the dayside magnetosphere size is higher for the CME with a southward negative $B_z$ component during the whole passage of the CME.

We also calculated the Ly-$\alpha$ transit absorption for all three CME cases. The transit spectra in the Ly-$\alpha$ line were calculated for the three different CME geometries and for three instants in each case: compressed magnetosphere ($t = 50$ min), enlarged magnetosphere ($t=100$ min), and relaxed magnetosphere ($t = 167$ min).
The transit absorption increases for all three cases of CME interaction compared to the pre-CME case. The radial CME field (Case-III) gives maximum blue-wing absorption of 8.7$\%$ (integrated over -230 km~s$^{-1}$ to -140 km~s$^{-1}$) during the peak CME interaction time.  This maximum absorption is smaller than the enhanced blue-wing absorption observed (14.4 $\pm$ 3.6 $\%$) for the HD189733b at September 2011 transit event \citep{Lecavelier12} and higher than the synthetic transit blue-wing absorption (5.1$\%$) presented in \citet{Hazra2022} using hydrodynamic simulations of CME-planet interaction. \citet{Rumenskikh2022} reported blue-wing Ly-$\alpha$ absorption of $\simeq 15\%$ for their hydrodynamic simulation of the CME interaction (strong stellar wind case) with the planetary atmosphere. For the strong stellar wind case, they found that the size of the region that produces the bulk of the Ly-$\alpha$ absorption extends out to $\sim$ 3R$_p$, similar to the extension of $\sim$ 2.5R$_p$ (maximum absorption area for the Case-III) found in our simulation during CME peak interaction. However, their larger absorption of $\simeq 15\%$ versus ours of 8.7$\%$ is likely due to the presence of ENAs, which contributed to half of the total absorption.

Within our parametric study, the interaction of a CME with the planetary magnetosphere plays a major role in the escape of planetary material.
If transit events could be observed while the CME is in interaction with the planet, an enhancement in the transit observation would be detected. This is most likely the case observed during the second transit event for the HD189733b \citep{Lecavelier12}. The magnetic field geometry of the embedded CME might not be purely northward, southward, or radial. They might have all components present in the CME. However, approximating them in simple $B_z$ geometries (both northward and southward) or $B_x$ geometry (radial only) gives us a reasonable understanding of how magnetic field plays a crucial role in enhancing the planetary mass loss and transit absorption.

\section*{Acknowledgements}
We thank an anonymous referee for their constructive comments that helped to improve the manuscript. GH acknowledges the IIT Kanpur Initiation Grant (IITK/PHY/2022386) for financial support. AAV, GH and SC acknowledge funding from the European Research Council (ERC) under the European Union’s Horizon 2020 research and innovation programme (grant agreement No 817540, ASTROFLOW). This work used the Dutch national e-infrastructure with the support of the SURF Cooperative using grant nos. EINF-2218 and EINF-5173. This work used the BATS-R-US tools developed at the University of Michigan Center for Space Environment Modeling and made available through the NASA Community Coordinated Modeling Center.

\section*{Data Availability}
The data underlying this article will be shared on reasonable request to the corresponding author.



\bibliographystyle{mnras}
\bibliography{myref_exo1, myref_exo2} 


\bsp	
\label{lastpage}
\end{document}